\documentclass{aa}

\usepackage{graphicx}
\usepackage{epsfig}

\newcommand{\pl}{\partial}
\renewcommand{\d}{{\rm d}}
\newcommand{\inta}{\int_{-i\infty}^{+i\infty}}
\newcommand{\beq}{\begin{equation}}
\newcommand{\eeq}{\end{equation}}
\newcommand{\beqa}{\begin{eqnarray}}
\newcommand{\eeqa}{\end{eqnarray}}
\newcommand{\bea}{\begin{array}}
\newcommand{\ea}{\end{array}}
\newcommand{\bx}{{\bf x}}
\newcommand{\bp}{{\bf p}}

\newcommand{\cG}{{\cal G}}
\newcommand{\df}{\delta f}
\newcommand{\rhob}{\overline{\rho}}
\newcommand{\bk}{{\bf k}}
\newcommand{\lag}{\langle}
\newcommand{\rag}{\rangle}
\newcommand{\bv}{{\bf v}}
\newcommand{\Om}{\Omega_{\rm m}}
\newcommand{\Ol}{\Omega_{\Lambda}}

\newcommand{\dLo}{\delta_{L0}}
\newcommand{\dR}{\delta_{R}}
\newcommand{\x}{{\vec \omega}}
\newcommand{\dL}{\delta_L}
\newcommand{\DL}{\Delta_L}
\newcommand{\DLo}{\Delta_{L0}}

\newcommand{\tDL}{\tilde{\Delta}_L}
\newcommand{\cF}{{\cal F}}

\newcommand{\Det}{{\rm Det}}

\newcommand{\cD}{{\cal D}}
\newcommand{\psib}{\overline{\psi}}
\newcommand{\cC}{{\cal C}}
\renewcommand{\Re}{{\rm Re}}
\renewcommand{\Im}{{\rm Im}}
\newcommand{\Arg}{{\rm Arg}}
\newcommand{\Cp}{{\cal C}_+}
\newcommand{\Cm}{{\cal C}_-}
\newcommand{\Ci}{{\cal C}_{\rm i}}
\newcommand{\Cy}{{\cal C}_y}

\newcommand{\Dgp}{D_+}

\newcommand{\cP}{{\cal P}}

\begin{document}

 

\title{Dynamics of gravitational clustering II. Steepest-descent method for the quasi-linear regime.}   
\author{P. Valageas}  
\institute{Service de Physique Th\'eorique, CEN Saclay, 91191 Gif-sur-Yvette, France} 
\date{Received / Accepted }

\abstract{
We develop a non-perturbative method to derive the probability distribution $\cP(\dR)$ of the density contrast within spherical cells in the quasi-linear regime. Indeed, since this corresponds to a rare-event limit a steepest-descent approximation can yield asymptotically exact results. We check that this is the case for Gaussian initial density fluctuations, where we recover most of the results obtained by perturbative methods from a hydrodynamical description. Moreover, we correct an error which was introduced in previous works for the high-density tail of the pdf. This feature, which appears for power-spectra with a slope $n<0$, points out the limitations of perturbative approaches which cannot describe the pdf $\cP(\dR)$ for $\dR \ga 3$ even in the limit $\sigma \rightarrow 0$. This break-up does not involve shell-crossing and it is naturally explained within our framework. Thus, our approach provides a rigorous treatment of the quasi-linear regime, which does not rely on the hydrodynamical approximation for the equations of motion. Besides, it is actually simpler and more intuitive than previous methods. Our approach can also be applied to non-Gaussian initial conditions.
}

\maketitle

\section{Introduction}

In standard cosmological scenarios large-scale structures in the universe arise from the growth of small initial density perturbations through gravitational instability, see \cite{Peebles1}. Besides, the amplitude of these density fluctuations usually increases at small scales, as in the CDM model (\cite{Peebles2}). This leads to a hierarchical scenario of structure formation where smaller scales become non-linear first. Then, at large scales or at early times one can use a perturbative approach to describe the evolution of the initial fluctuations. This is usually done through an hydrodynamical description (e.g., \cite{Fry1}, \cite{Goroff1}). Thus, one describes the dark matter as a pressure-less fluid which obeys the continuity and Euler equations, coupled with the Poisson equation for the gravitational potential. However, as soon as shell-crossing appears this hydrodynamical description becomes inexact and one can no longer associate only one velocity to each spatial position. This implies that the perturbative series must diverge for hierarchical scenarios (with no small-scale cutoff). Nonetheless, the perturbative results obtained by both hydrodynamical and Boltzmann approaches are actually identical (e.g., \cite{paper1}).

The disadvantage of such recursive perturbative procedures, where one computes in serial order the successive terms of the perturbative expansion, is that they can only be used for the first few order terms (e.g., up to order 3). Indeed, the calculations become rather heavy for high-order terms. Hence this method cannot be used to estimate the high-order cumulants of the probability distribution of the density field $\rho(\bx)$, since the cumulant of order $q$ depends on the term of order $(q-1)$ of the perturbative expansion. Nevertheless, for the case of Gaussian initial conditions it has been shown that one could use the structure of the perturbative expansion to obtain at leading order in the limit $\sigma \rightarrow 0$ all cumulants of any order of the density contrast $\dR$ within spherical cells (Bernardeau 1992,1994), where $\sigma$ is the rms density fluctuation. This allows one to get the precise shape of the probability distribution function (pdf) $\cP(\dR)$ in the quasi-linear regime. However, this derivation presents several shortcomings. First, it is based on the perturbative expansion of the density field while this series actually diverges for hierarchical scenarios (e.g., \cite{paper1}). Hence the proof of the results obtained by this perturbative method is not complete. Second, it does not apply to non-Gaussian primordial density fluctuations.

In this article, we present a non-perturbative method to obtain the pdf $\cP(\dR)$ of the density contrast in the quasi-linear regime. It is based on a steepest-descent approximation which yields exact results in the asymptotic limit $\sigma \rightarrow 0$. Thus, it provides a rigorous justification of most of the previous perturbative calculations and it allows us to correct an error introduced in those works. Besides, it is actually much more intuitive. Another advantage of our approach is that we can also study non-Gaussian primordial density fluctuations, as we discuss in a companion paper (\cite{paper3}).

This article is organized as follows. In Sect.\ref{Functional formulation} we recall the equations of motion and we introduce the generating functions which describe the statistical properties of the density field. Then, in Sect.\ref{Saddle-point method} we describe the steepest-descent method which allows us to derive the pdf $\cP(\dR)$ in the quasi-linear regime for Gaussian initial conditions. We also present convenient geometrical constructions of the relevant generating function. Finally, in Sect.\ref{Comparison with previous results} we compare our method with previous results published in the literature.

\section{Functional formulation}
\label{Functional formulation}

\subsection{Equations of motion}
\label{Equations of motion}

The gravitational dynamics of a collisionless fluid is described by the collisionless Boltzmann equation coupled with the Poisson equation. Since we consider in this article the quasi-linear regime it is convenient to use the comoving coordinate $\bx$. Then, we define the impulsion $\bp$ by:
\beq
\bp = a^2 \dot{\bx}
\label{p1}
\eeq
where $a(t)$ is the scale factor and we note $f(\bx,\bp,t)$ the distribution function. Thus, $f(\bx,\bp,t) \d^3x \d^3p$ is the mass enclosed in the phase-space element $\d^3x \d^3p$. Next, we define the perturbation $\df$ of the distribution function by:
\beq
f(\bx,\bp,t) = \rhob \; \delta_D(\bp) + \rhob \; \df(\bx,\bp,t)
\label{df1}
\eeq
where $\rhob$ is the mean comoving density (it is constant with time) and $\delta_D$ is Dirac's function. Then, the density contrast $\delta(\bx,t)$ is simply given by:
\beq
\delta(\bx,t) = \int \df(\bx,\bp,t) \; \d\bp .
\label{delta1}
\eeq
Finally, we define the spatial Fourier transform of the field $\df$ by:
\beq
\left\{ \bea{l} {\displaystyle \df(\bx,\bp,t) = \int \d\bk \; e^{i \bk.\bx} \; \df(\bk,\bp,t) }  \\ \\ {\displaystyle \df(\bk,\bp,t) = \frac{1}{(2\pi)^3} \int \d\bx \; e^{-i \bk.\bx} \; \df(\bx,\bp,t) } \ea \right.
\label{Four1}
\eeq
Then, as shown in \cite{paper1} the collisionless Boltzmann equation can be written:
\beqa
\frac{\pl \df}{\pl t} + i \frac{\bk.\bp}{a^2} \df + i \frac{4\pi \cG \rhob}{a}  \frac{\bk}{k^2} . \frac{\pl \delta_D}{\pl \bp}(\bp) \int \d\bp' \; \df(\bk,\bp') \nonumber \\  \hspace{0.cm} + i \frac{4\pi \cG \rhob}{a} \int \d\bk' \d\bp' \df(\bk',\bp') \frac{\bk'}{k'^2} . \frac{\pl \df}{\pl \bp}(\bk-\bk',\bp) = 0
\label{Bol1}
\eeqa
after we used the Poisson equation to substitute for the gravitational potential $\phi$. Thus, the distribution function $\df$ is fully determined by eq.(\ref{Bol1}) supplemented with initial conditions. As shown in \cite{paper1}, these initial conditions can be defined by the linear growing mode $\eta(\bk,\bp,t)$ which is a solution of the linearized eq.(\ref{Bol1}). Moreover, this linear solution $\eta(\bk,\bp,t)$ can be derived from the linear mode of the hydrodynamical equations and we have:
\beq
\eta = D_+ \dLo(\bk) \delta_D(\bp) - i \; a^2 \dot{D}_+ \dLo(\bk) \frac{\bk}{k^2} . \frac{\pl \delta_D}{\pl \bp}(\bp)
\label{eta1}
\eeq
where $D_+(t)$ is the usual linear growing mode and $\dLo(\bk)$ is the linear density contrast today (i.e. at redshift $z=0$). In \cite{paper1} we developed a method to obtain the solution of eq.(\ref{Bol1}) as a perturbative expansion over the linear mode $\eta$. In particular, we explained that perturbative results obtained from the collisionless Boltzmann equation are identical to those derived from the hydrodynamical approach. However, these perturbative series are only asymptotic (i.e. the series actually diverge, see \cite{paper5}). Nevertheless, the key point is that the initial conditions can be defined by the linear mode $\eta(\bk,\bp,t)$ even if the distribution function $\df$ cannot be written as a perturbative series over $\eta$, as shown in \cite{paper1}.

\subsection{Functional $Z[j]$}
\label{Functional}

In order to describe gravitational clustering in the universe we do not need to obtain the explicit solution of eq.(\ref{Bol1}) for all possible initial conditions. Indeed, since the linear mode $\eta$ which sets the initial conditions is a random field we are only interested in the statistical properties of the distribution function $\df$. These are fully described by the functional $Z[j]$ of the test field $j(\bx,\bp,t)$ defined by:
\beq
Z[j] \equiv  \lag e^{\int \d\bx \d\bp \d t \; j . \df} \rag
\label{Z1}
\eeq
where $\lag .. \rag$ expresses the average over the initial conditions. If we expand the exponential in eq.(\ref{Z1}) we can also write:
\beqa
Z[j] & = & 1 + \sum_{q=1}^{\infty} \frac{1}{q!} \int \d\x_1 .. \d\x_q\; j(\x_1) .. j(\x_q)  \nonumber \\ & & \times \; \lag \df(\x_1) .. \df(\x_q) \rag
\eeqa
where we noted $\x$ the 7-dimensional coordinate $\x=(\bx,\bp,t)$. This expression clearly shows that the determination of the functional $Z[j]$ is equivalent to the derivation of multi-time correlation functions at all orders. As seen in eq.(\ref{eta1}) the initial conditions $\eta(\bx,\bp,t)$ are fully defined by the linear density contrast today $\dLo(\bk)$ since they can be restricted to the linear growing mode. Moreover, we assume in this article that these density fluctuations $\dLo(\bk)$ are Gaussian. This is consistent with usual scenarios of structure formation based on the simplest inflationary models, e.g. \cite{Bar1}. Then, the average in eq.(\ref{Z1}) is given by a Gaussian weight over the random field $\dLo(\bk)$. In real space $\bx$ we obtain the path-integral:
\beq
Z[j] = \left( \Det\DLo^{-1} \right)^{1/2}  \int [\d\dLo(\bx)] \; e^{j.\df -\frac{1}{2} \dLo . \DLo^{-1} . \dLo} .
\label{Z2}
\eeq
The normalization factor ensures that $Z[0]=1$, as implied by the definition (\ref{Z1}) (and $\Det\DLo^{-1}$ is the determinant of the kernel $\DLo^{-1}$). Here $\df$ has to be understood as the distribution function which is the solution of the equation of motion (\ref{Bol1}) determined by the initial condition $\eta(\x)$ defined by $\dLo(\bk)$ as in eq.(\ref{eta1}). We also used the short-hand notation $j.\df \equiv \int \d\x \; j(\x) . \df(\x)$ and $\dLo . \DLo^{-1} . \dLo \equiv \int \d\bx_1\d\bx_2 \; \dLo(\bx_1) . \DLo^{-1}(\bx_1,\bx_2) . \dLo(\bx_2)$. Note that the fields $j$ and $\df$ depend on the 7-dimensional coordinate $(\bx,\bp,t)$ while $\dLo$ only depends on the spatial coordinate $\bx$. The kernel (i.e. infinite dimensional matrix) $\DLo^{-1}$ is the inverse of the kernel:
\beq
\DLo(\bx_1,\bx_2) \equiv \lag \dLo(\bx_1) \dLo(\bx_2) \rag
\label{Dlo1}
\eeq
which fully defines the statistics of the random field $\dLo(\bx)$ since the latter is Gaussian. We define the Fourier transform of the kernel $\DLo$ by the property:
\beq
f_1 . \DLo . f_2 = \int \d\bk_1 \d\bk_2 \; f_1(\bk_1)^{\ast} . \DLo(\bk_1,\bk_2) . f_2(\bk_2)
\label{FourD}
\eeq
for any real fields $f_1$ and $f_2$. Using eq.(\ref{Four1}) this implies:
\beq
\DLo(\bk_1,\bk_2) = \int \d\bx_1 \d\bx_2 \; e^{i(\bk_2.\bx_2 - \bk_1.\bx_1)} \; \DLo(\bx_1,\bx_2)
\label{KerFour1}
\eeq
which gives:
\beq
\DLo(\bk_1,\bk_2) = (2\pi)^6 \; P_0(k_1) \; \delta_D(\bk_1 - \bk_2)
\label{Dlo2}
\eeq
where we defined the power-spectrum $P_0(k)$ of the linear density contrast today by:
\beq
\lag \dLo(\bk_1) \dLo(\bk_2) \rag \equiv P_0(k_1) \; \delta_D(\bk_1+\bk_2) .
\label{P01}
\eeq
Then, since the inverse $\DLo^{-1}$ is defined by the property:
\beq
\int \d\bx \; \DLo^{-1}(\bx_1,\bx) \; \DLo(\bx,\bx_2) = \delta_D(\bx_1-\bx_2)
\label{invertx}
\eeq
which reads in Fourier space:
\beq
\int \d\bk \; \DLo^{-1}(\bk_1,\bk) \; \DLo(\bk,\bk_2) = (2\pi)^6 \; \delta_D(\bk_1-\bk_2)
\label{invert}
\eeq
we can see that the kernel $\DLo$ obtained in eq.(\ref{Dlo2}) can be inverted as:
\beq
\DLo^{-1}(\bk_1,\bk_2) = \frac{1}{P_0(k_1)} \; \delta_D(\bk_1-\bk_2) .
\label{Dlo3}
\eeq
Moreover, we can see from eq.(\ref{FourD}) and eq.(\ref{Dlo3}) that the inverse kernel $\DLo^{-1}$ is positive definite since we have:
\beq
\dL . \DLo^{-1} . \dL = \int \d\bk \; \frac{|\dL(\bk)|^2}{P_0(k)}
\label{Dlo4}
\eeq
where we used $\dL(-\bk)=\dL(\bk)^{\ast}$ for real fields $\dL(\bx)$, see eq.(\ref{Four1}). Note that $\DLo(\bx_1,\bx_2)$ is simply the two-point correlation function $\xi_{L0}(|\bx_2-\bx_1|)$ of the linear density field today. Hence $\DLo(\bx_1,\bx_2)$ is symmetric, homogeneous and isotropic, as implied by eq.(\ref{Dlo2}).

Note that in \cite{paper1} we obtained an alternative path-integral expression for the functional $Z[j]$. It involved an integration over the actual distribution $\df(\x)$ and the associated weight was not Gaussian: the argument of the exponential contained terms of order two to four over the field $\df$. Moreover, all the terms were explicitly known. By contrast, the path-integral (\ref{Z2}) involves a simple Gaussian integration over the initial conditions but the factor $j.\df$ is not explicitly known. Nevertheless, the expression (\ref{Z2}) will prove to be more convenient because we shall not need the explicit mapping $\df[\dLo]$. Indeed, as shown in the next sections we shall only need particular spherical solutions.

\subsection{Generating functions}
\label{Generating functions}

In Sect.\ref{Functional} we introduced the functional $Z[j]$ which provides all statistical properties of the stochastic field $\df(\bx,\bp,t)$. However, in practice one does not need all of these properties of $\df$. In particular, one is often mainly interested in the pdf $\cP(\dR)$ of the density contrast $\dR$ within a spherical cell of comoving radius $R$, volume $V$:
\beq
\dR = \int_V \frac{\d^3x}{V} \; \delta(\bx) .
\label{dR}
\eeq
Without any loss of generality we can take this cell $V$ to be centered on the origin $\bx=0$. Next, it is convenient to introduce the generating function $\psi(y)$ defined by:
\beq
\psi(y) \equiv \lag e^{-y \dR} \rag \equiv \int_{-1}^{\infty} \d\dR \; e^{-y \dR} \; \cP(\dR)  .
\label{psi1}
\eeq
Here $\lag .. \rag$ is the average over the initial conditions. The last equality in eq.(\ref{psi1}) defines the pdf $\cP(\dR)$. Note that the generating function $\psi_L(y)$ associated with the pdf $\cP_L(\delta_{L,R})$ of the linear density contrast is related to $\cP_L$ as in eq.(\ref{psi1}) where the integration over $\delta_{L,R}$ now runs from $-\infty$ to $+\infty$. The pdf can be recovered from $\psi(y)$ through the inverse Laplace transform:
\beq
\cP(\dR) = \inta \frac{\d y}{2\pi i} \; e^{y \dR} \; \psi(y) .
\label{P1}
\eeq
In fact, as shown by eq.(\ref{psi1}) one can directly obtain the moments of the pdf from the generating function $\psi(y)$:
\beq
\psi(y) = \sum_{q=0}^{\infty} \frac{(-y)^{q}}{q!} \; \lag \dR^q \rag .
\label{psiseries1}
\eeq
Comparing eq.(\ref{psi1}) with eq.(\ref{Z1}) we see that we have $\psi(y) = Z[j_R]$ with:
\beq
j_R(\bx,\bp,t) = - y \; \frac{\theta(x<R)}{V} \; \delta_D(t-t_R)
\label{jR}
\eeq
where $\theta(x<R)$ is a top-hat with obvious notations and $t_R$ is the time which we consider. Then, instead of integrating over the linear density contrast today $\dLo$ as in eq.(\ref{Z2}) we can equivalently integrate over the density contrast at the time of interest $\dL$ (since $\dL \propto \dLo$). This yields:
\beq
\psi(y) = \left( \Det\DL^{-1} \right)^{1/2} \int [\d\dL(\bx)] \; e^{-y \dR[\dL] -\frac{1}{2} \dL . \DL^{-1} . \dL}
\label{psi2}
\eeq
where we introduced the symmetric kernel:
\beq
\DL(\bx_1,\bx_2) \equiv \lag \dL(\bx_1) \dL(\bx_2) \rag = \xi_L(|\bx_2-\bx_1|) .
\label{DL1}
\eeq
Thus, $\DL^{-1}$ is the inverse operator of the two-point correlation function $\xi_L$ of the linear density field. In eq.(\ref{psi2}) the quantity $\dR[\dL]$ is the exact non-linear density contrast $\dR$ over the cell $V$ which arises from the gravitational dynamics of the linear density field $\dL(\bx)$, as described by the equation of motion (\ref{Bol1}). Note that eq.(\ref{FourD}) to eq.(\ref{Dlo4}) also apply to $\DL$, $\DL^{-1}$ and $P(k)$.

\section{Steepest-descent method}
\label{Saddle-point method}

\subsection{Action $S[\dL]$}
\label{Action}

The calculation of path-integrals such as (\ref{psi2}) is in general a rather difficult task. However, when a parameter becomes very small one may try a steepest-descent approximation. Indeed, it may happen that in such a limit the integral in eq.(\ref{psi2}) becomes increasingly dominated by the point where the argument of the exponential is maximum (i.e. the minimum of the ``action''). See for instance any textbook on Quantum Field Theory for a discussion of the steepest-descent approximation. In this article we consider the quasi-linear regime. Then, the parameter which tends to zero is the amplitude of the linear two-point correlation $\DL$, that is the amplitude of the linear power-spectrum $P(k)$ at the time of interest.

In order to factorize the amplitude of the two-point correlation $\DL$ it is convenient to define a new generating function $\varphi(y,\sigma)$ by:
\beq
\psi(y) \equiv e^{-\varphi(y \sigma^2,\sigma)/\sigma^2}
\label{psisig}
\eeq
where we noted as usual $\sigma(R)$ the rms linear density fluctuation in a cell of radius $R$:
\beq
\sigma^2(R) \equiv \lag \delta_{L,R}^2 \rag = \int_V \frac{\d\bx_1}{V} \frac{\d\bx_2}{V} \; \DL(\bx_1,\bx_2) .
\label{xibL}
\eeq
Here $\delta_{L,R}$ is the linear density contrast within the spherical cell of radius $R$. Hereafter, we consider the shape of the power-spectrum to be fixed and the second argument of $\varphi(y,\sigma)$ describes the dependence of the pdf on the amplitude of the power-spectrum. The generating function $\varphi(y,\sigma)$ provides the pdf $\cP(\dR)$ through eq.(\ref{P1}). It also yields the cumulants $\lag \dR^q \rag_c$ of the density contrast through:
\beq
\varphi(y,\sigma) = \sum_{q=2}^{\infty} \frac{(-1)^{q-1}}{q!} \; \frac{\lag \dR^q \rag_c}{\sigma^{2(q-1)}} \; y^q
\label{cum1}
\eeq
where we used the fact that $\lag \dR \rag=0$. Using eq.(\ref{psi2}) we obtain:
\beq
e^{-\varphi(y,\sigma)/\sigma^2} = \left( \Det\DL^{-1} \right)^{1/2} \int [\d\dL(\bx)] \; e^{- S /\sigma^2(R)}
\label{phi1}
\eeq
where we introduced the action $S$:
\beq
S[\dL] \equiv y \; \dR[\dL] + \frac{\sigma^2(R)}{2} \; \dL . \DL^{-1} . \dL
\label{S0}
\eeq
The action $S[\dL]$ is independent of the normalization of the linear power-spectrum $P(k)$ since $\DL \propto \sigma^2$. Then, it is clear that the path-integral in eq.(\ref{phi1}) is dominated by the minimum of the action $S$ in the limit $\sigma \rightarrow 0$ for a fixed $y$. Indeed, the contributions from other states $\dL$ are exponentially damped relative to this point. Moreover, the steepest-descent approximation becomes exact in this limit. Of course, this is the reason why we performed the change of variable $y \rightarrow y/\sigma^2$ in the change $\psi \rightarrow \varphi$ in eq.(\ref{psisig}).

\subsection{Spherical saddle-point}
\label{Spherical saddle-point}

Thus, for any $y$ we look for the point $\dL$ where the action $S$ is minimum. The condition which expresses that $\dL$ is an extremum (or a saddle-point) is:
\beq
\frac{\delta S}{\delta(\dL(\bx))} = 0 \hspace{0.3cm} \mbox{for all} \hspace{0.3cm} \bx ,
\label{extremum}
\eeq
where $\delta/\delta(\dL(\bx))$ is the functional derivative with respect to $\dL$ at the point $\bx$. This constraint also writes:
\beq
y \; \frac{\delta(\dR)}{\delta(\dL(\bx))} + \sigma^2(R) \int \d\bx' \; \DL^{-1}(\bx,\bx') \dL(\bx') = 0 
\label{S2}
\eeq
since $\DL^{-1}$ is symmetric. Hereafter, we consider $y$ to be real. Multiplying both sides of eq.(\ref{S2}) by the operator $\DL/\sigma^2(R)$ we get after a change of notations (see eq.(\ref{invertx}) for the property which characterizes inverse operators):
\beq
\dL(\bx') = \frac{- y}{\sigma^2(R)} \int \d\bx'' \; \DL(\bx',\bx'') \; \frac{\delta(\dR)}{\delta(\dL(\bx''))} .
\label{S3}
\eeq
Since $\DL(\bx_1,\bx_2)= \xi_L(|\bx_2-\bx_1|)$ is homogeneous and isotropic we can look for a solution of eq.(\ref{S3}) which is spherically symmetric. For such a spherical solution, $\dL(\bx')$ is fully defined by the overall density contrast $\delta_{L,R'}$ within spherical cells of arbitrary radius $R'$ and volume $V'$ centered on the origin. Then, taking the mean of eq.(\ref{S3}) over a cell $V'$ we get:
\beq
\delta_{L,R'} = - y \int_{V'} \frac{\d\bx'}{V'} \int \d\bx'' \; \frac{\DL(\bx',\bx'')}{\sigma^2(R)} \; \frac{\delta(\dR)}{\delta(\dL(\bx''))} .
\label{S4}
\eeq
Going back to the definition of derivatives we can write eq.(\ref{S4}) as:
\beqa
\delta_{L,R'} & = & \lim_{\epsilon \rightarrow 0} \frac{-y}{\epsilon} \biggl \lbrace \dR \left[ \dL(\bx'') + \epsilon \int_{V'} \frac{\d\bx'}{V'} \; \frac{\DL(\bx',\bx'')}{\sigma^2(R)} \right] \nonumber \\ & & \hspace{1.5cm} - \; \dR \left[ \dL(\bx'') \right] \biggl \rbrace .
\label{S5}
\eeqa
We can see that both arguments of the functionals $\dR$ which appear in eq.(\ref{S5}) are spherically symmetric. Hence we only need consider the restriction of $\dR[\dL(\bx'')]$ to spherically symmetric states, which we note $\dR[\delta_{L,R''}]$. Here $\bx''$ and $R''$ are dummy variables. Thus, we write eq.(\ref{S5}) as:
\beqa
\delta_{L,R'} = \lim_{\epsilon \rightarrow 0} \frac{-y}{\epsilon} \left\{ \dR \left[ \delta_{L,R''} + \epsilon \tDL(R',R'') \right] - \dR \left[ \delta_{L,R''} \right] \right\}  \nonumber \\
\label{col1}
\eeqa
where we introduced the normalized kernel:
\beqa
\lefteqn{ \tDL(R_1,R_2) \equiv \frac{\lag \delta_{L,R_1} \delta_{L,R_2} \rag}{\sigma^2(R)} } \nonumber \\ & = & \int_{V_1} \frac{\d\bx_1}{V_1} \int_{V_2} \frac{\d\bx_2}{V_2} \; \frac{\DL(\bx_1,\bx_2)}{\sigma^2(R)} 
\label{tD1}
\eeqa
which is independent of the normalization of the power-spectrum and it satisfies: $\tDL(R,R)=1$. Thus, the linear density profile $\delta_{L,R''}$ must satisfy the constraint (\ref{col1}) for any radius $R'$. This fully defines the saddle-point $\delta_{L,R''}$. The matter enclosed within the radius $R$ for a spherical linear state $\delta_{L,R''}$ comes from a Lagrangian comoving radius $R_L$. Moreover, if there is no shell-crossing the actual density contrast $\dR$ only depends on the linear density contrast $\delta_{L,R_L}$ because of the spherical symmetry of the system. Thus, for such spherically symmetric states we can write:
\beq
\left\{ \begin{array}{l}
{\displaystyle \dR = \cF \left[ \delta_{L,R_L} \right] } \\ \\
{\displaystyle R_L^3 = (1+\dR) R^3 }
\end{array} \right.
\label{F1}
\eeq
where the function $\cF[\delta_{L,R_L}]$ is given by the usual spherical collapse solution of the equations of motion. For instance, in the case of a critical density universe $\cF(\dL)$ is defined by the implicit systems (\cite{Peebles1}):
\beq 
\delta_L \geq 0 \; : \;\; \left\{ \begin{array}{rl}
1+\cF(\dL) & = {\displaystyle \frac{9}{2} \; \frac{(\theta-\sin
\theta)^2}{(1-\cos \theta)^3} } \\ \\ \delta_L & = {\displaystyle
\frac{3}{20} \; \left[ 6 (\theta-\sin \theta) \right]^{2/3} } \end{array}
\right.
\label{Fcol1}
\eeq 
and 
\beq 
\delta_L < 0 \; : \;\; \left\{ \begin{array}{rl}
1+\cF(\dL) & = {\displaystyle \frac{9}{2} \; \frac{(\sinh
\eta-\eta)^2}{(\cosh \eta-1)^3} } \\ \\ \delta_L & = {\displaystyle -
\frac{3}{20} \; \left[ 6 (\sinh \eta-\eta) \right]^{2/3} } \end{array}
\right.
\label{Fcol2}
\eeq 
The second equation in (\ref{F1}) merely expresses the conservation of mass. In order to solve eq.(\ref{col1}) we need the change of $\dR$ to first order in $\epsilon$ under the transformation:
\beq
\delta_{L,R''} \rightarrow \delta_{L,R''} + \epsilon \; \tDL(R',R'') .
\label{eps1}
\eeq
Under this infinitesimal transformation both quantities $\dR$ and $R_L$ are modified and we write to first order in $\epsilon$:
\beq
\dR \rightarrow \dR + \epsilon \; \delta_R^{(1)} \hspace{0.3cm} \mbox{and} \hspace{0.3cm} R_L \rightarrow R_L + \epsilon \; R_L^{(1)} .
\eeq
Substituting into the system (\ref{F1}) we obtain for the first-order term:
\beq
\left\{ \begin{array}{l}
{\displaystyle \delta_R^{(1)} = \cF' \left[ \delta_{L,R_L} \right] \; \left( R_L^{(1)} \left. \frac{\d \delta_{L,R''}}{\d R''} \right|_{R_L} + \tDL(R',R_L) \right) } \\ \\
{\displaystyle 3 R_L^2 R_L^{(1)} = \delta_R^{(1)} R^3 }
\end{array} \right.
\label{F2}
\eeq
which yields $\delta_R^{(1)}$. From eq.(\ref{col1}) we see that $\delta_{L,R'} = - y \delta_R^{(1)}$ and we obtain:
\beq
\delta_{L,R'} = - y \; \frac{ \cF' \left[ \delta_{L,R_L} \right] \tDL(R',R_L) }{ 1 - \cF' \left[ \delta_{L,R_L} \right] \frac{R^3}{3R_L^2} \left. \frac{\d \delta_{L,R''}}{\d R''}\right|_{R_L} } .
\label{col2}
\eeq
The only term in the r.h.s. of eq.(\ref{col2}) which depends on $R'$ is the factor $\tDL(R',R_L)$ hence we have:
\beq
\delta_{L,R'} = \delta_{L,R_L} \; \frac{\tDL(R',R_L)}{\tDL(R_L,R_L)} .
\label{col3}
\eeq
As a consequence, we can write:
\beq
\frac{\d \delta_{L,R''}}{\d R''} = \delta_{L,R_L} \; \frac{1}{\tDL(R_L,R_L)} \frac{\pl \tDL}{\pl R''}(R'',R_L) .
\label{der1}
\eeq
Here we note that we also have the identity:
\beq
\sigma^2(R'') = \sigma^2(R) \;{\tDL(R'',R'')} .
\label{sig1}
\eeq
Then, since $\tDL(R_1,R_2)$ is symmetric over $R_1 \leftrightarrow R_2$ we can easily see from eq.(\ref{sig1}) that:
\beq
\frac{1}{\tDL(R_L,R_L)} \; \left. \frac{\pl \tDL}{\pl R''}(R'',R_L) \right|_{R_L} = \frac{1}{\sigma(R_L)} \; \frac{\d \sigma}{\d R}(R_L) .
\label{sig2}
\eeq
Hence, using eq.(\ref{der1}) we can write eq.(\ref{col2}) at the point $R'=R_L$ as:
\beq
\delta_{L,R_L} = - y \; \frac{ \cF' \left[ \delta_{L,R_L} \right] \sigma^2(R_L)/\sigma^2(R) }{ 1 - \cF' \left[ \delta_{L,R_L} \right] \frac{R^3}{3R_L^2} \delta_{L,R_L} \frac{1}{\sigma(R_L)} \; \frac{\d \sigma}{\d R}(R_L) } .
\label{col4}
\eeq
This fully defines the spherically symmetric saddle-point $\dL(\bx)$. The implicit equation (\ref{col4}) determines $\delta_{L,R_L}$ while the radial profile of this initial state is given by eq.(\ref{col3}). The radius $R_L$ is related to $R$ by the second equation of (\ref{F1}). Note that this saddle-point is an exact solution of the collisionless Boltzmann equation: we do not use an hydrodynamical description of the system.

\subsection{Density profile of the saddle-point}
\label{Density profile}

Before we derive the generating function $\varphi(y,\sigma)$ implied by the spherical saddle-point obtained in the previous section, we need to examine its radial density profile. For instance, if the density contrast were to become larger than unity at radii greater than $R_L$ this would invalidate the previous results since these outer shells would have collapsed and relations (\ref{F1}) would no longer hold. The radial profile is given by eq.(\ref{col3}). Let us define the Fourier transform $F(kR)$ of the top-hat of radius $R$ by:
\beq
F(kR) \equiv \int_V \frac{\d\bx}{V} \; e^{i \bk . \bx} = 3 \; \frac{\sin(kR)-(kR)\cos(kR)}{(kR)^3} .
\label{window1}
\eeq
Then, we obtain from eq.(\ref{tD1}) and eq.(\ref{P01}):
\beq
\tDL(R_1,R_2) = \frac{1}{\sigma^2(R)} \int \d\bk \; F(kR_1) \; F(kR_2) \; P(k)
\eeq
which yields for the density profile:
\beq
\delta_{L,R'} = \delta_{L,R_L} \; \frac{ \int \d\bk \; F(kR') \; F(kR_L) \; P(k) } {\int \d\bk \; F(kR_L)^2 \; P(k)} .
\label{profil1}
\eeq

\begin{figure}
{\epsfxsize=8cm {\epsfbox{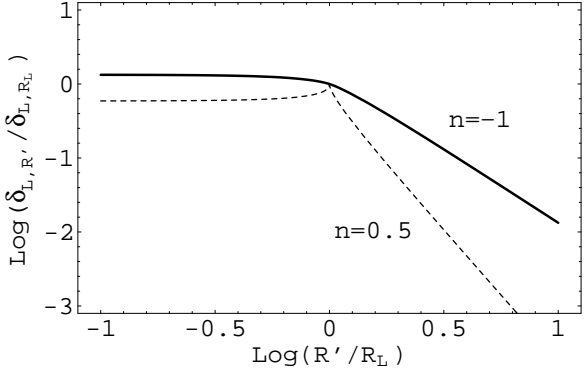}} }
\caption{Cumulative linear density profile $\delta_{L,R'} / \delta_{L,R_L}$ of the spherical saddle-point. The solid line corresponds to $n=-1$ and the dashed-line to $n=0.5$.} 
\label{figprofil}
\end{figure}

To get an idea of the radial profile implied by eq.(\ref{profil1}) it is convenient to consider the case of a power-law linear power-spectrum $P(k) \propto k^n$. Then, we can write eq.(\ref{profil1}) as:
\beq
\frac{\delta_{L,R'}}{\delta_{L,R_L}} = \left( \frac{R'}{R_L} \right)^{-3/2} \frac{ \int_0^{\infty} \d k \; k^{n-1} J_{3/2}(kR')  J_{3/2}(kR_L) } { \int_0^{\infty} \d k \; k^{n-1} \;  J_{3/2}(kR_L)^2 }
\label{profil2}
\eeq
where we used the fact that, using eq.(\ref{window1}), $F(x)$ can also be written:
\beq
F(x) = 3 \; \sqrt{\frac{\pi}{2}} \; x^{-3/2} \; J_{3/2}(x) .
\label{window2}
\eeq
Here $J_{3/2}(x)$ is the standard Bessel function of order $3/2$. The expression (\ref{profil2}) also reads:
\beq
\frac{\delta_{L,R'}}{\delta_{L,R_L}} = \left( \frac{2 R_L}{R_L+R'} \right)^{n+3} \; \frac{ _2 F_1 \left( \frac{n+3}{2} , 2 ; 4 ; \frac{4 R_L R'}{(R_L+R')^2} \right) } { _2 F_1 \left( \frac{n+3}{2} , 2 ; 4 ; 1 \right) }
\label{window3}
\eeq
where $_2 F_1$ is Gauss' Hypergeometric function, see \cite{Grad1} (\S 6.576.2). Then, using the relations (see \cite{Grad1}, \S 9.122.1):
\beq
_2 F_1 \left( \frac{n+3}{2} , 2 ; 4 ; 1 \right) = \frac{24}{(1-n)(3-n)}
\eeq
and
\beq
_2 F_1 \left( \frac{n+3}{2} , 2 ; 4 ; 0 \right) = 1 ,
\eeq
we obtain the asymptotic expressions:
\beq
R' \rightarrow 0 : \;\; \frac{\delta_{L,R'}}{\delta_{L,R_L}} \rightarrow 2^n \; \frac{(1-n)(3-n)}{3}
\label{profil3}
\eeq
and
\beq
R' \rightarrow \infty : \; \frac{\delta_{L,R'}}{\delta_{L,R_L}} \sim 2^n \; \frac{(1-n)(3-n)}{3}  \left( \frac{R_L}{R'} \right)^{n+3} .
\label{profil4}
\eeq
Note that for all cases of cosmological interest we have $-3<n<1$. Thus, we see that at small radii $\delta_{L,R'}$ remains finite, of the order of $\delta_{L,R_L}$, while at large radii it decreases as $R'^{-(n+3)} \propto \sigma^2(R')$. Therefore, the derivation of the spherical saddle-point in Sect.\ref{Spherical saddle-point} is valid since for $|\delta_{L,R_L}| \ll 1$ the density contrast remains small at all radii so that eq.(\ref{F1}) and eq.(\ref{col2}) hold for all radii $R'$. For illustrative purposes we show in Fig.\ref{figprofil} the cumulative linear density profile of the spherical saddle-point for the cases $n=-1$ and $n=0.5$. Note that for $n>0$ the amplitude of the density perturbation actually shows a slow decline for $R'<R_L$. Moreover, for $n>0$ the local density contrast $\dL(\bx)$ changes sign at $|\bx|=R_L$ (so that $|\delta_{L,R'}|$ decreases faster than $R'^{-3}$ at large radii) while for $n=0$ the profile $\dL(\bx)$ is simply a top-hat of radius $R_L$.

\subsection{Generating function $\varphi(y)$}
\label{Generating function}

The spherical saddle-point we obtained in the previous section provides the asymptotic behaviour of the pdf $\cP(\dR)$ in the limit $\sigma \rightarrow 0$. Thus, it yields the limiting generating function $\varphi(y)$ defined by:
\beq
\varphi(y) \equiv \varphi(y,\sigma=0) .
\label{phisig0}
\eeq
To write $\varphi(y)$ we first need the value $S_y$ of the action $S[\dL]$ at the saddle-point, which is given by eq.(\ref{S0}). Since the saddle-point obtained in Sect.\ref{Spherical saddle-point} is spherically symmetric we can check from eq.(\ref{col3}) and the definition (\ref{tD1}) that it obeys:
\beq
\dL(\bx) =  \delta_{L,R_L} \; \frac{1}{\tDL(R_L,R_L)} \int_{V_L} \frac{\d\bx'}{V_L} \; \frac{\DL(\bx,\bx')}{\sigma^2(R)} .
\label{col5}
\eeq
Substituting this expression into eq.(\ref{S0}) we obtain:
\beq
S_y = y \; \cF\left[\delta_{L,R_L}\right] + \frac{1}{2} \; \frac{\sigma^2(R)}{\sigma^2(R_L)} \; \delta_{L,R_L}^2 .
\label{S01}
\eeq
Then, applying the steepest-descent method we approximate the path-integral in eq.(\ref{phi1}) by the Gaussian integration around the spherical saddle-point. This yields:
\beq
e^{-\varphi(y,\sigma)/\sigma^2} = \left( \Det\DL^{-1} \right)^{1/2} \; \left( \Det M_y \right)^{-1/2} \; e^{- S_y /\sigma^2(R)}
\label{phi2}
\eeq
where the minimum $S_y$ of the action $S$ is given by eq.(\ref{S01}) while the matrix $M_y$ is the Hessian of the exponent:
\beqa
\lefteqn{ M_y(\bx_1,\bx_2) \equiv \frac{\delta^2 (S/\sigma^2)}{\delta(\dL(\bx_1)) \delta(\dL(\bx_2))} } \nonumber \\ & & = \frac{y}{\sigma^2(R)} \; \frac{\delta^2 (\dR)}{\delta(\dL(\bx_1)) \delta(\dL(\bx_2))} +  \DL^{-1}(\bx_1,\bx_2) . 
\label{M1}
\eeqa
Note that eq.(\ref{phi2}) becomes exact in the limit $\sigma \rightarrow 0$ if the spherical saddle-point found in the previous sections is indeed the global minimum of the action. Taking the logarithm of eq.(\ref{phi2}) we see that the contributions to $\varphi(y)$ from the determinants vanish in the limit $\sigma \rightarrow 0$ so that we simply get:
\beq
\varphi(y) = S_y = y \; \cF\left[\delta_{L,R_L}\right] + \frac{\tau(\delta_{L,R_L})^2}{2}
\label{tau1}
\eeq
where we introduced the function $\tau(\dL)$ defined by:
\beq
\tau(\dL) \equiv \frac{- \; \dL \; \sigma(R)}{\sigma \left[ (1+\cF[\dL])^{1/3} R \right] } .
\label{tau2}
\eeq
We introduced a minus sign in eq.(\ref{tau2}) in order to recover the results of the perturbative hydrodynamical approach (see eq.(\ref{G2}) and eq.(\ref{tau4}) below). Next, taking the derivative $\tau'(\dL)$ we can recognize the structure of the denominator in eq.(\ref{col4}) and we can write eq.(\ref{col4}) as:
\beq
\tau \; \frac{\d\tau}{\d\dL} = - y \; \cF'(\dL) .
\label{tau3}
\eeq
Thus, the intersection of the curves $\tau \d\tau/\d\dL$ and $- y \cF'(\dL)$ determines the saddle-point $\delta_{L,R_L}$, or $\tau$, associated with a given value of $y$. We can simplify this implicit system by introducing the function $\cG(\tau)$:
\beq
\cG(\tau) \equiv \cF\left[\dL(\tau)\right] = \dR
\label{G1}
\eeq
where $\dL(\tau)$ is defined by eq.(\ref{tau2}). Using eq.(\ref{tau2}) we can see that $\cG(\tau)$ is also defined by the implicit relation:
\beq
\cG(\tau) = \cF\left[ - \tau \; \frac{\sigma\left[(1+\cG[\tau])^{1/3}R\right]}{\sigma(R)} \right] .
\label{G2}
\eeq
Then, eq.(\ref{tau3}) and eq.(\ref{tau1}) can be written:
\beq
\left\{ \begin{array}{l}
{\displaystyle \tau = - y \; \cG'(\tau) } \\ \\
{\displaystyle \varphi(y) = y \; \cG(\tau) + \frac{\tau^2}{2} }
\end{array} \right.
\label{tau4}
\eeq
This implicit system fully defines the generating function $\varphi(y)$, with $\cG(\tau)$ defined in eq.(\ref{G2}). As discussed in Sect.\ref{Comparison with previous results}, this agrees with the results obtained from the usual perturbative hydrodynamical approach (Bernardeau (1992, 1994)). Note that for a power-law power-spectrum $P(k) \propto k^n$ we have $\sigma(R) \propto R^{-(n+3)/2}$ so that eq.(\ref{G2}) simplifies to:
\beq
\cG(\tau) = \cF\left[ - \tau \; (1+\cG[\tau])^{-(n+3)/6} \right] .
\label{G3}
\eeq
Finally, we note that the definition (\ref{Fcol1}) for the function $\cF(\dL)$ breaks down for $\dL \ga 1$ where shell-crossing occurs. In fact, beyond this point the density contrast $\dR$ also depends on the density profile of the initial condition so that we can no longer write a relation of the form (\ref{F1}). Hence, our results only hold for $\dL \la 1$ (i.e. as long as there is no shell-crossing beyond the radius $R$). However, in the quasi-linear regime $\sigma \ll 1$ the typical values of $\dL$ are small: $|\dL| \sim \sigma(R)$. As seen from eq.(\ref{tau2}) and (\ref{tau4}) this corresponds to $|y| \sim |\tau| \sim \sigma(R) \ll 1$. Thus, in the quasi-linear regime the pdf $\cP(\dR)$ should be well-described by the system (\ref{tau4}). Indeed, this limit corresponds to $\sigma \rightarrow 0$ for a fixed $y$, which we can take to be small but finite (e.g., $|y| \la 0.1$).

However, before we can reach this conclusion we must check two points. First, we must make sure that the saddle-point $\dL$ we obtained in Sect.\ref{Spherical saddle-point} is indeed a minimum of the action (and not a maximum for instance). Second, we must ensure that it is the global minimum (and not a mere local minimum). 

To check the first point we simply need to make sure that the Hessian $W(\bx_1,\bx_2)$ of the action $S[\dL]$ is positive definite at this point. Since $W=\sigma^2(R) M_y$ where $M_y$ is the kernel defined in eq.(\ref{M1}) we have:
\beqa
\lefteqn{ W(\bx_1,\bx_2) \equiv \frac{\delta^2 S}{\delta(\dL(\bx_1)) \delta(\dL(\bx_2))} } \label{Wdef} \\ & & = y \; \frac{\delta^2 (\dR)}{\delta(\dL(\bx_1)) \delta(\dL(\bx_2))} + \sigma^2(R) \; \DL^{-1}(\bx_1,\bx_2) . 
\label{W1}
\eeqa
The kernel $\sigma^2(R) \; \DL^{-1}(\bx_1,\bx_2)$ is of order unity since it is independent of the normalization of the power-spectrum and of $y$. Moreover, it is positive definite, as shown by eq.(\ref{Dlo4}). On the other hand, the first term in the r.h.s. of eq.(\ref{W1}) vanishes as $y$ for $y \rightarrow 0$ (indeed, $\delta_{L,R_L} \rightarrow 0$ for $y\rightarrow 0$ so that the second derivative tends to its value at the point $\dL=0$). Then, if we define the determinant $D(y) \equiv \mbox{det}(W)$ we have $D(0)>0$. Since $D(y)$ is a continuous function of $y$ we can conclude that $D(y)>0$ over a finite range of $y$ around 0, which implies that $W$ remains positive definite over a finite range of $y$. Hence, for small $y$ the saddle-point we obtained in Sect.\ref{Spherical saddle-point} is indeed a minimum of the action $S$.

Next, we must show that this local minimum is in fact a global minimum of the action. As explained above, we take $y$ to be small (but finite), $|y| \ll 1$, since we study the quasi-linear regime. On the other hand, the second term in eq.(\ref{W1}) is of order unity. Then, we see that if there exists another local minimum $\dL'$ of $S$ it must be at least of order unity. Indeed, in the neighbourhood of the spherical saddle-point $\dL$ where the Hessian $W$ is dominated by $\sigma^2 \DL^{-1}$ there can be no other saddle-point.

Let us first consider the case of positive $y$. As seen from eq.(\ref{tau4}) and eq.(\ref{tau2}) this corresponds to positive $\tau$ and negative $\dL$ (since $\cG(\tau)$ is a decreasing function of $\tau$). In fact, this could be directly seen from eq.(\ref{S0}) which clearly shows that in order to minimize the action $S$ with a positive $y$ we must have $\dR<0$. Moreover, since the density $\rho$ must be positive we have the constraint $\dR \geq -1$. Hence we obtain from eq.(\ref{S0}):
\beq
y \geq 0 : \; S[\dL] \geq - y + \frac{\sigma^2(R)}{2} \; \dL . \DL^{-1} . \dL
\label{S0m}
\eeq
Since $|y| \ll 1$ and $\dL' \sim 1$ we see from eq.(\ref{S0m}) that $S[\dL'] \simeq (\sigma^2(R)/2) \; (\dL' . \DL^{-1} . \dL') \ga 1$ since $\DL^{-1}$ is positive definite. On the other hand, we can check from eq.(\ref{tau1}) that $S[\dL] \simeq -y^2/2 <0$ for the spherical saddle-point we obtained in Sect.\ref{Spherical saddle-point} (indeed, for $|y| \ll 1$ we have: $\tau \simeq y \simeq -\delta_{L,R_L} \simeq -\dR$). Hence if there exists another minimum $\dL'$ of $S$ it obeys $S[\dL'] > S[\dL]$. Thus, we conclude that for small positive $y$ the saddle-point obtained in Sect.\ref{Spherical saddle-point} is the global minimum of the action, which justifies the previous calculation.

Finally, we consider the case $y<0$, which corresponds to $\dR>0$. This case is more difficult since there is no upper bound for $\dR$ and for large $\dL$ we no longer have a relation of the form (\ref{F1}). In fact, we shall see below that for a linear power-spectrum with $n<0$ the saddle-point $\dL$ is not the global minimum of the action. Actually, in this case the action is no longer bounded from below. Then, the steepest-descent method described above is a priori no longer justified. In fact, a specific study shows that it is still useful but it requires some care. We shall come back to this point in the next section.

\subsection{Geometrical construction}
\label{Geometrical construction}

In order to get an intuitive picture of the generating function defined by the system (\ref{tau4}) it is convenient to devise a geometrical construction which yields $\tau(y)$ and $\varphi(y)$. First, we note that the first line of eq.(\ref{tau4}) simply states that the implicit function $\tau(y)$ is given by the intersection of the straight line $\tau/y$ of variable slope $1/y$ with the fixed generating function $-\cG'(\tau)$.

\begin{figure}[htb]
\centerline{\epsfxsize=8cm \epsfysize=5.5cm \epsfbox{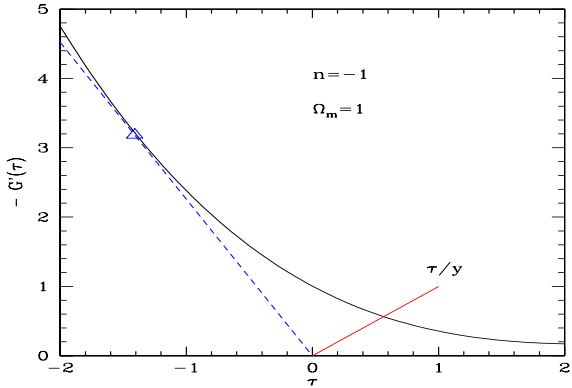}}
\caption{Construction of the function $\tau(y)$ for a linear power-spectrum with $n=-1$ in a critical density universe. To each value of $y$ is associated the abscissa $\tau$ of the intersection of the straight line $\tau/y$ with the curve $-\cG'(\tau)$. The dashed-line with the triangle shows the location of the singularity $\tau_s \simeq -1.4$, $y_s\simeq -0.44$.} 
\label{figGtaud1}
\end{figure}

This construction is shown in Fig.\ref{figGtaud1} for the case of a linear power-spectrum with $n=-1$ in a critical density universe. From eq.(\ref{G3}) we obtain the inverse $\tau(\cG)$ as:
\beq
\tau(\cG) = - (1+\cG)^{(n+3)/6} \; (1+\cF)^{-1}[1+\cG]
\label{G4}
\eeq
where $(1+\cF)^{-1}$ is the inverse of the function $1+\cF$. Then, using eq.(\ref{Fcol1}) and eq.(\ref{Fcol2}), keeping in mind that $\cG$ is the actual density contrast (eq.(\ref{G1})), we get from eq.(\ref{G4}) the asymptotic behaviours:
\beq
\mbox{high densities} : \; \tau \rightarrow -\infty , \; \cG \rightarrow \infty : \; \cG \sim (-\tau)^{6/(n+3)}
\label{tauG1}
\eeq
and:
\beq
\mbox{low densities} : \; \tau \rightarrow \infty , \; \cG \rightarrow -1 : \; (1+\cG) \sim \tau^{-6/(1-n)}
\label{tauG2}
\eeq
where we did not write positive numerical multiplicative factors of order unity. Thus, for large negative $\tau$ we have $-\cG' \sim (-\tau)^{(3-n)/(3+n)}$ which increases faster than $|\tau|$ for $n<0$. This implies that there is a minimum value $y_s<0$ of $y$ for which the straight line $\tau/y$ can intersect the curve $-\cG'(\tau)$. This is shown by the dashed-line in Fig.\ref{figGtaud1}. On the other hand, for $y_s<y<0$ we see that we have two intersection points $\tau_- <\tau_s<\tau_+<0$. Thus the function $\tau(y)$ is bivaluate over this range.

\begin{figure}[htb]
\centerline{\epsfxsize=8cm \epsfysize=5.5cm \epsfbox{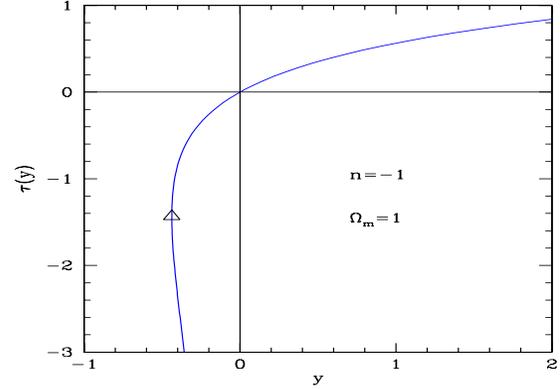}}
\caption{The function $\tau(y)$ for a linear power-spectrum with $n=-1$ in a critical density universe. The triangle shows the location of the singularity $\tau_s \simeq -1.4$, $y_s\simeq -0.44$.} 
\label{figtauy}
\end{figure}

We display in Fig.\ref{figtauy} the function $\tau(y)$ we obtain for the case shown in Fig.\ref{figGtaud1}. Note that the function $y(\tau)$ is well-behaved and shows no singularity. The singularity $(\tau_s,y_s)$ is given by the point where $\d\tau/\d y=\infty$. From eq.(\ref{tau4}) this condition also reads:
\beq
\cG'(\tau_s) = \tau_s \cG''(\tau_s) , \;\; y_s = - \frac{\tau_s}{\cG'(\tau_s)} = - \frac{1}{\cG''(\tau_s)} .
\label{sing1}
\eeq

The advantage of the geometrical construction displayed in Fig.\ref{figGtaud1} is that it shows at once the location of possible singular points. However, there is an alternative geometrical construction which directly yields the value of the generating function $\varphi(y)$. Indeed, we can also write the full system (\ref{tau4}) as the one equation:
\beq
\varphi(y) = \min_{\tau} \left[ S_y(\tau) \right] \;\;\; \mbox{with} \;\;\; S_y(\tau) = y \; \cG(\tau) + \frac{\tau^2}{2}  .
\label{phi5}
\eeq
Of course, the minimum which appears in eq.(\ref{phi5}) expresses the fact that the saddle-point we look for is the minimum of the action $S[\dL]$. Note that eq.(\ref{phi5}) means that we minimize the action $S$ over the subspace of spherical linear states $\dL$ of the form (\ref{col5}) parameterized by $\tau$ (or $\delta_{L,R_L}$). However, we must actually minimize the action over all possible states $\dL$. The rigorous justification of eq.(\ref{phi5}) was obtained in Sect.\ref{Spherical saddle-point} and Sect.\ref{Generating function} where we showed that the spherical minimum described by eq.(\ref{phi5}) is also a minimum with respect to transverse directions. Then, we can give the following geometrical solution of eq.(\ref{phi5}).

\begin{figure}[htb]
\centerline{\epsfxsize=8cm \epsfysize=5.8cm \epsfbox{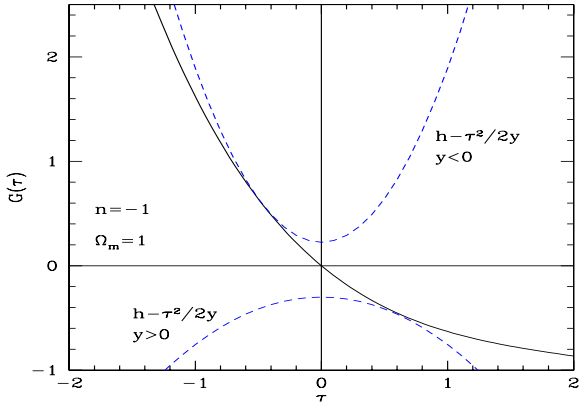}}
\caption{Construction of the function $\varphi(y)$ for a linear power-spectrum with $n=-1$ in a critical density universe. The quantity $\varphi(y)/y$ is simply the height $h$ of the parabola $h-\tau^2/(2y)$ at the first contact with the curve $\cG(\tau)$. For $y>0$ (i.e. underdensities) we start from below at $h=-\infty$ while for $y<0$ (i.e. overdensities) we start from above at $h=+\infty$.} 
\label{figGtau}
\end{figure}

First, for $y>0$ we note that the point $\tau$ where $S_y(\tau)$ is minimum is also the point where $S_y(\tau)/y = \cG(\tau) + \tau^2/(2y)$ is minimum. Then, this point is simply given by the first contact of the parabola $h-\tau^2/(2y)$, of varying height $h$, with the curve $\cG(\tau)$, starting from below at $h=-\infty$. Then, the minimum of the action is given by $\varphi(y)=y \times h$ at this point. Second, for $y<0$ we need the maximum of $S_y(\tau)/y$. This is given by the first contact of the parabola $h-\tau^2/(2y)$ with the curve $\cG(\tau)$, starting from above at $h=+\infty$, and we have again $\varphi(y)=y \times h$. This construction is displayed in Fig.\ref{figGtau}. In particular, it is clear that for small $y$ the parabola are very narrow and we get only one contact point at $\tau \sim y$ as we probe the small-$\tau$ part of the curve $\cG(\tau)$ where $\cG(\tau) \simeq -\tau$. That is the curvature of the parabola gets very large with $y \rightarrow 0$ while the curvature of $\cG(\tau)$ is finite. This is the essence of the discussion below eq.(\ref{W1}). Thus, this geometrical construction gives at once the value of the generating function $\varphi(y)$. In particular, one can see at a glance from the curve $\cG(\tau)$ the behaviour of $\varphi(y)$. Note that if there exists a singular point $(\tau_s,y_s)$, as in Fig.\ref{figGtau}, the minimum obtained for small negative $y$ is only a local minimum. We shall come back to this point in Sect.\ref{calPDF}.

\begin{figure}[htb]
\centerline{\epsfxsize=8cm \epsfysize=5.8cm \epsfbox{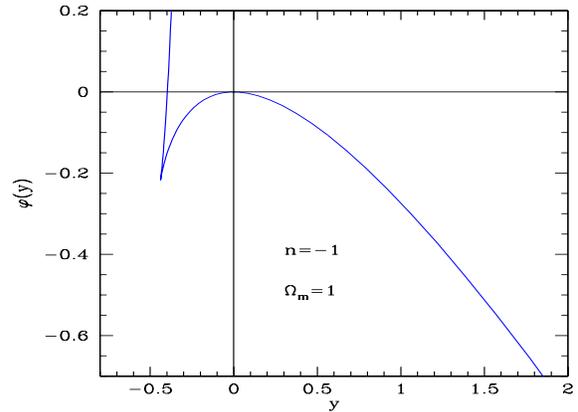}}
\caption{The generating function $\varphi(y)$ for a linear power-spectrum with $n=-1$ in a critical density universe. The feature at $y=y_s \simeq -0.44$ clearly displays the singularity of $\varphi(y)$ at this point. This ``regularized'' generating function $\varphi(y)$ exhibits a regular branch at the origin and it is bivaluate over the range $y_s<y<0$.} 
\label{figphi}
\end{figure}

Finally, we display in Fig.\ref{figphi} the generating function $\varphi(y)$ obtained for this same case $n=-1$ (and $\Om=1$). The feature at $y=y_s$ shows that $\varphi(y)$ is singular at this point. The curve drawn in Fig.\ref{figphi} is obtained from the parametric system (\ref{tau4}) with $-\infty<\tau<\infty$. As explained from Fig.\ref{figGtaud1} the function $\tau(y)$ is bivaluate over $y_s<y<0$. This also applies to $\varphi(y)$. The branch which runs through the origin in Fig.\ref{figphi} corresponds to $\tau>\tau_s$ while the upper branch over $y_s<y<0$ which starts almost vertically at $y_s$ corresponds to $\tau<\tau_s$.

\subsection{Calculation of the pdf $\cP(\dR)$}
\label{calPDF}

The generating function $\varphi(y)$ was obtained in Sect.\ref{Generating function} for real values of $y$, using a steepest-descent method. Then, using eq.(\ref{P1}) and eq.(\ref{psisig}) we obtain the pdf $\cP(\dR)$ through the inverse Laplace transform:
\beq
\cP(\dR) = \inta \frac{\d y}{2\pi i \sigma^2(R)} \; e^{[y \dR - \varphi(y)]/\sigma^2(R)} .
\label{P2}
\eeq
Next, in order to compute numerically the pdf $\cP(\dR)$ through eq.(\ref{P2}) we need to continue the function $\varphi(y)$ over the complex plane. This is simply done by using the same implicit system (\ref{tau4}) for complex values of $y$. Note that the function $\cG(\tau)$ is analytic over the region of interest. Indeed, the function $\cF(\dL)$ defined in eq.(\ref{Fcol1}) and eq.(\ref{Fcol2}) is analytic, with singular points such that $\cos \theta=1$ or $\cosh \eta =1$ (with $\theta \neq 0$ and $\eta \neq 0$). However, the singularity at $\dL \simeq 1.68$ (i.e. for $\theta=2\pi$) is repelled to $\tau = -\infty$, as shown in eq.(\ref{G4}) and eq.(\ref{tauG1}). Hence the function $\cG(\tau)$ has no singularities along the real axis.

Then, we need to specify the integration path over $y$ in eq.(\ref{P2}). It intersects the real axis at the saddle-point $(\tau_c,y_c)$ given by:
\beq
\frac{\d\chi}{\d y}(y_c) =0 \hspace{0.2cm} \mbox{with} \hspace{0.2cm} \chi(y) \equiv y \; \dR - \varphi(y) .
\label{Chic1}
\eeq
From eq.(\ref{tau4}) we have:
\beq
\cG(\tau_c) = \dR  \hspace{0.2cm} \mbox{since} \hspace{0.2cm} \varphi'(y) = \cG(\tau) .
\label{tauc}
\eeq
Thus, we see from eq.(\ref{G1}) that this triplet $(\dR,\tau_c,y_c)$ is also the triplet $(\dR,\tau,y)$ we obtained in Sect.\ref{Spherical saddle-point} and Sect.\ref{Generating function} to get $\varphi(y)$. Of course, this is required by self-consistency. It simply means that $\cP(\dR)$ at the point $\dR$ is governed by the neighbourhood of the saddle-point $\dL(\bx)$ obtained in Sect.\ref{Spherical saddle-point}, which obeys $\cF(\delta_{L,R_L})=\dR$. This result agrees with intuitive expectations. Then, the integration path in the complex plane is set by the constraint Im$(\chi)=0$ with the requirement that Re$(\chi)$ decreases on both sides from its value at $y_c$. This corresponds to the steepest-descent path. The second derivative of the factor $\chi$ is:
\beq
\chi''(y) = - \varphi''(y) .
\eeq
As seen in Fig.\ref{figphi} we have $\varphi''(y) <0$ (if there is no singularity $y_s$). In fact, for $|y| \ll 1$ we get: $\varphi''(y) \simeq -1$. Therefore, the path of steepest-descent is orthogonal to the real axis at the point $y_c$. Moreover, it is symmetric about the real axis, which clearly shows from eq.(\ref{P2}) that the result for $\cP(\dR)$ is real, since $\varphi(y^{\ast})=\varphi(y)^{\ast}$. Thus, we can write eq.(\ref{P2}) as:
\beq
\cP(\dR) = \Im \int_{y_c}^{i\infty} \frac{\d y}{\pi \sigma^2(R)} \; e^{[y \dR - \varphi(y)]/\sigma^2(R)}
\label{P3}
\eeq
where we only integrate over the upper half-plane $\Im(y) \geq 0$. Note that in the quasi-linear limit $\sigma \rightarrow 0$ the contribution to the integral (\ref{P2}) only comes from an infinitesimal neighbourhood of the saddle-point $(\tau_c,y_c)$ around the real axis. Therefore, we could try a steepest-descent approximation for eq.(\ref{P2}). However, the agreement with the results of numerical simulations is better if we numerically compute the exact integral (\ref{P2}). Thus, in the following we compute the full integral (\ref{P2}). 

Note that this means that it is better to approximate the generating function $\varphi(y)$, and next to use the exact inverse Laplace transform (\ref{P2}), rather than to directly approximate the pdf $\cP(\dR)$. This can be understood as follows. In the limit $\sigma \rightarrow 0$ a steepest-descent approximation to eq.(\ref{P2}) would be fully justified (it is actually exact in this limit). However, it is clear that if we use the results obtained for $\cP(\dR)$ in the limit $\sigma \rightarrow 0$ (that is, we assume that we have obtained in some way the behaviour of $\cP(\dR)$ at all points $\dR$ at leading-order in this limit) for a finite value $\sigma >0$ we can generically expect that the moments $\lag \dR^q \rag$ obtained from this approximate pdf exhibit sub-leading terms which are not correct. In particular, this means that we would get: $\lag 1 \rag = 1 + o(1)$ and $\lag \dR \rag = o(1)$ where $o(1)$ stands for a term which vanishes in the limit $\sigma \rightarrow 0$. Generically, we may expect this term to be of order unity when $\sigma \sim 1$. This implies that for small but finite $\sigma$ the pdf is not exactly normalized to unity and the mean $\lag \dR \rag$ is not exactly zero. By contrast, using the exact inverse Laplace transform (\ref{P2}) with the generating function $\varphi(y)$ (obtained in the limit $\sigma \rightarrow 0$) ensures that we have for any finite $\sigma$ the exact integrals $\lag 1 \rag =1$ and $\lag \dR \rag = 0$. Thus, the normalization and the mean are always correct. This result can be obtained from the expansion (\ref{psiseries1}) and eq.(\ref{psisig}) which shows that in order to have the exact moments of order 0 and 1 we only need $\varphi(y)$ to be quadratic in $y$ for $y \rightarrow 0$. Of course, this is the case since from eq.(\ref{tau4}) we have the expansion $\varphi(y)=-y^2/2 + ...$. On the other hand, eq.(\ref{P2}) implies that $\lag \dR^2 \rag = \sigma^2$ for any $\sigma$.

The procedure we described above allows us to compute the pdf $\cP(\dR)$ in the quasi-linear regime, using the steepest-descent method developed in the previous sections. However, when the function $\cG(\tau)$ grows faster than $\tau^2$ as $\tau \rightarrow -\infty$ a singularity $y_s$ shows up in the generating function $\varphi(y)$ and matters are slightly more involved. First, we note that for such functions $\cG(\tau)$, which corresponds to $n<0$ as shown by eq.(\ref{tauG1}), there is no global minimum of the action $S[\dL]$ for negative $y$. This is clear from the construction of Fig.\ref{figGtau}. Indeed, it is obvious that the contact point shown in Fig.\ref{figGtau} at $\tau \simeq -0.5$ for the upper parabola is only a local minimum and there is no global minimum: whatever large $h$ is taken to be, the parabola always intersects the curve $\cG(\tau)$. This is also clear from eq.(\ref{phi5}). Indeed, we now get $S_y(\tau) \rightarrow -\infty$ for $\tau \rightarrow -\infty$. Hence the ``action'' $S_y(\tau)$ is not bounded from below if $y<0$. This actually means that the path-integrals (\ref{psi2}) and (\ref{phi1}) diverge for $y<0$. Hence the generating functions $\psi(y)$ and $\varphi(y)$ exhibit a branch cut along the negative real axis. Then, the steepest-descent method described in the previous sections must be modified (in fact, there may still exist a global minimum if we take into account shell-crossing, which appears for large $\dL$ or large negative $\tau$, but this is irrelevant here). Note that a negative $\dR$ corresponds to positive $\tau_c$ and $y_c$, as shown by eq.(\ref{tauc}) and Fig.\ref{figGtau}. Hence this problems only appears when one looks for the value of the pdf $\cP(\dR)$ for positive $\dR$.

We can note that from a physical point of view the pdf $\cP(\dR)$ at the point $\dR$ should still be governed by the saddle-point $(\tau_c,y_c)$ obtained in the previous sections. Indeed, it is clear that a non-linear density contrast $\dR$ arises from initial conditions close to the spherical saddle-point derived in Sect.\ref{Spherical saddle-point}. In fact, there is a straightforward trick to show this in a more explicit fashion. Indeed, as we explained above the problem is due to the rapid growth of the functional $\dR[\dL]$ for large positive $\dL$ (we do not consider shell-crossing here since it is not related to this problem). Then, instead of looking for the pdf $\cP(\dR)$ we can as well investigate the pdf $\cP(\dR^{1/q})$ where $q$ is a large odd integer. Obviously, the steepest-descent method developed in Sect.\ref{Spherical saddle-point} can be applied to this new pdf. This involves new generating functions $\psi_q(y) \equiv \lag e^{-y \dR^{1/q}} \rag$ and $\varphi_q(y)$. We again obtain a spherical saddle-point of the form (\ref{col5}) and the implicit system (\ref{tau4}) where the new function $\cG_q(\tau)$ is simply: $\cG_q(\tau) = \cG(\tau)^{1/q}$. Note that the saddle-point $\delta_{L,R_L}$ associated with a given non-linear density contrast $\dR$ does not depend on $q$. Of course, this was to be expected since to a given $\dR$ corresponds a well-defined set of initial states $\dL(\bx)$, whatever we consider $\dR$ itself or $\dR^{1/q}$! Then, we see that if we choose a large enough value for $q$ the function $\cG_q(\tau)$ grows more slowly than $\tau^2$ for $\tau \rightarrow -\infty$. Therefore, we can now apply the steepest-descent method as described in the previous section. Note that the new generating function $\varphi_q(y)$ shows no singularity $y_{s,q}$ so that we span the whole curve $\cG_q(\tau)$ (hence $\cG(\tau)$). Finally, from $\cP(\dR^{1/q})$ we can derive $\cP(\dR)$ through a simple change of variables. For all $q$ we obtain in this way the same exponential-like cutoff (i.e. the exponential of a given power of the density) at large densities but the multiplicative factor obtained in the limit $\sigma \rightarrow 0$ will usually differ. In other words, in order to get a unique and well-defined result we must take into account the determinants which appear in eq.(\ref{phi2}): we have to keep $\sigma$ small but finite.

\begin{figure}[htb]
\centerline{\epsfxsize=8cm \epsfysize=5.8cm \epsfbox{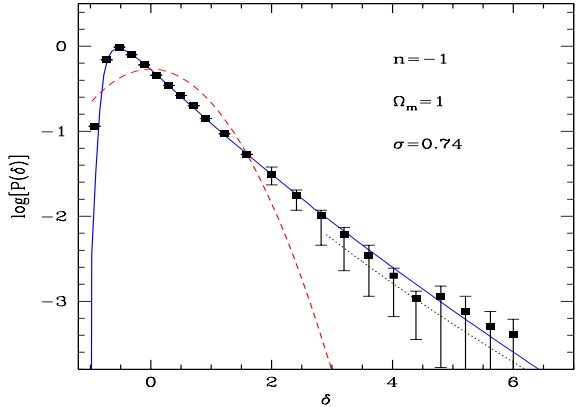}}
\caption{The pdf $\cP(\dR)$ for $n=-1$, $\Om=1$ and $\sigma=0.74$. The data points (obtained from numerical simulations) are taken from \cite{Ber2}. The solid line shows the theoretical prediction from eq.(\ref{P2}) and eq.(\ref{tau4}). The dotted line shows the contribution from the unstable saddle-point $\tau_-$ which gives the high-density tail of the pdf, from eq.(\ref{PdRdisc2}). The dashed-line displays the Gaussian with the same variance.} 
\label{figPdelta}
\end{figure}

In fact, as described in App.\ref{A worked example} and App.\ref{Application} we can directly work with the density contrast $\dR$, even though the path-integral (\ref{psi2}) diverges for negative real $y$. One must simply be careful to use appropriate integration contours in the complex plane when using integrals like (\ref{P2}). Thus, the integration paths we use in eq.(\ref{P2}) are shown in Fig.\ref{figCy1}. They depend on the density contrast $\dR$ and they run through the spherical saddle-point derived in Sect.\ref{Spherical saddle-point}. In particular, in agreement with the simple procedure described above based on $\dR^{1/q}$, for large $\dR$ the pdf is governed by the saddle-point $(\tau_c,y_c)$ given by eq.(\ref{tauc}). This means that for high density contrasts we span the upper branch of $\varphi(y)$ shown in Fig.\ref{figphi}. Since the action $S[\dL]$ is not bounded from below for $y_s<y<0$ the two saddle-points $\tau_-$ and $\tau_+$ (with $\tau_- < \tau_s < \tau_+<0$) obtained in Sect.\ref{Geometrical construction} are not the global minimum of the action (which does not exist). The saddle-point $\tau_+$ (which corresponds to the branch of $\varphi(y)$ which runs through the origin in Fig.\ref{figphi}) is only a local minimum. On the other hand, the point $\tau_-$ (the upper branch of $\varphi(y)$) is an unstable saddle-point: it is a local maximum of the action. However, as shown in App.\ref{A worked example} and App.\ref{Application} the Laplace transform $\psi(y)$ and the pdf $\cP(\dR)$ are still governed by these saddle-points in the quasi-linear regime. In particular, the saddle-point $\tau_-$ yields the high-density tail of the pdf. Note moreover that the generating function $\varphi(y)$ obtained in eq.(\ref{tau4}) and shown in Fig.\ref{figphi} is not the exact generating function. Indeed, as we noticed above the actual generating function shows a branch cut on the real negative axis (i.e. for $y<0$ and not only $y<y_s$). This is also explained in App.\ref{A worked example}.

Finally, we show in Fig.\ref{figPdelta} the pdf $\cP(\dR)$ obtained for $\sigma=0.74$ and $n=-1$ in the quasi-linear limit. The solid line is the result obtained from eq.(\ref{P2}) and eq.(\ref{tau4}). That is, we use the branch which runs through the origin of the generating function defined by eq.(\ref{tau4}) and displayed in Fig.\ref{figphi}. This curve was already obtained in \cite{Ber2} through a perturbative method (see also Sect.\ref{Perturbative methods}). Since this function exhibits a branch cut on the real axis at $y<y_s$ it yields an exponential high-density tail of the form $\cP(\dR) \sim e^{y_s \dR/\sigma^2(R)}$ (note $y_s<0$) as can be seen from eq.(\ref{P2}) (the integral is dominated by $y \simeq y_s$), see also App.\ref{A worked example}. This describes the pdf $\cP(\dR)$ for $\dR < \cG(\tau_s)$ in the quasi-linear limit. However, for larger density contrasts the pdf is governed by the unstable saddle-point $\tau_-$, which also led to the upper branch of the ``regularized'' $\varphi(y)$ in Fig.\ref{figphi}. Then, as shown in App.\ref{A worked example}, by closing the integration contour onto the negative real axis the inverse Laplace transform (\ref{P2}) can be written as:
\beq
\cP(\dR) = - \int_{-\infty}^0 \frac{\d y}{\pi \sigma^2} \; e^{y \dR/\sigma^2} \; \Im [ \psib(y) ] .
\label{PdRdisc1}
\eeq
Here we noted $2i\;\Im[\psib(y)]$ the discontinuity of the generating function $\psib(y) \equiv \psi(y/\sigma^2) = e^{-\varphi(y)/\sigma^2}$ along the negative real axis, see also eq.(\ref{psilog5}) and eq.(\ref{Plog3}). We compute the pdf $\cP(\dR)$ in App.\ref{Application} through a steepest-descent approximation, which yields:
\beq
\cP(\dR) \simeq \frac{1}{\sqrt{2\pi}\sigma} \; \frac{1}{1+\dR} \; \frac{1}{|\cG'(\tau)|} \; e^{-\tau^2/(2\sigma^2(R))} .
\label{PdRdisc2}
\eeq
The pdf obtained in this way is shown by the dotted-line in Fig.\ref{figPdelta}. It yields a smoother cutoff for the high-density tail of the pdf. Indeed, using the asymptotic form (\ref{tauG1}) we get from eq.(\ref{PdRdisc2}):
\beq
\dR \gg 1 : \;\; \cP(\dR) \sim e^{-\dR^{(n+3)/3}/\sigma^2(R)}
\label{tailP1}
\eeq
where we did not write positive multiplicative factors of order unity in the exponent. Thus, we see that for $n<0$, where the singular point $(\tau_s,y_s)$ appears, the pdf exhibits a high-density cutoff which is shallower than a pure exponential. Of course, this clearly implies that the Laplace transform $\psi(y)$ defined in eq.(\ref{psi1}) shows a branch cut on the real negative axis since the integral in the r.h.s. of eq.(\ref{psi1}) must diverge for $y<0$. This is at the origin of the additional difficulties encountered in the case $n<0$, where an artificial singularity $y_s$ appears. This is discussed in App.\ref{A worked example}. The form (\ref{tailP1}) is actually valid for all $n$ (i.e. also for $n>0$ where the analysis is simpler since the relevant saddle-point $\tau$ is really the global minimum of the action). It explicitly shows that the high-density cutoff of the pdf depends on the initial conditions (the slope $n$ of the linear power-spectrum). This is actually quite natural and it agrees with a simple intuitive spherical model (\cite{Val1}), as discussed below in Sect.\ref{Comparison with previous results}. We note in Fig.\ref{figPdelta} that for $\sigma=0.74$ the unstable saddle-point contribution only dominates $\cP(\dR)$ for $\dR \ga 8$ which is larger than $\cG(\tau_s) \simeq 3$. This is due to the finite value of $\sigma$. Thus, in the limit $\sigma \rightarrow 0$, as soon as $\dR>\cG(\tau_s)$ this contribution is larger than the result one would get by using for $\varphi(y)$ only the branch which runs through the origin in Fig.\ref{figphi}. Here, we must note that for large density contrasts the form (\ref{PdRdisc2}) is no longer valid since shell-crossing comes into play. Then, one must take into account virialization processes. However, we shall not study these very high densities here since we restrict ourselves to the quasi-linear regime where such events are extremely rare. Finally, we can see in Fig.\ref{figPdelta} that the quasi-linear limit provides a very good estimate of $\cP(\dR)$ up to $\sigma \sim 1$, which is a rather large value. Note also the strong departure of the pdf from the Gaussian (shown as the dashed-line).

\section{Comparison with previous results}
\label{Comparison with previous results}

\subsection{Perturbative methods}
\label{Perturbative methods}

Eventually, we point out that the results we obtained in Sect.\ref{Saddle-point method} partly agree with the standard results derived from a perturbative hydrodynamical approach. Indeed, the system (\ref{tau4}) which gives the generating function $\varphi(y)$ was also obtained by \cite{Ber2}. This result was derived from a perturbative expansion of the density field over the linear growing mode, substituted into the equations of motion of the hydrodynamical description. Note that our calculation does not involve the hydrodynamical approximation: it is based on the collisionless Boltzmann equation. However, as explained in \cite{paper1} the perturbative expansions obtained in both approaches actually coincide. Hence it is not surprising that we recover most of the results of \cite{Ber2}. 

On the other hand, we stress that the method we presented in this article is actually {\it non-perturbative}. In particular, we did not need to assume that the density field can be written as a perturbative expansion. This is important since as explained in \cite{paper1} and \cite{paper5} this perturbative expansion actually diverges (it is only asymptotic). Moreover, our calculation directly provides the pdf $\cP(\dR)$ in Eulerian space and we do not need to go from Lagrangian space to Eulerian space, which is a delicate step in the usual method. In particular, there is no need to apply any ``smoothing'' a posteriori: we directly obtain the pdf of the density field at a given scale $R$ which enters into the formulation of the problem itself. Thus, our calculation provides the needed justification of these previous results. For instance, if the spherical saddle-point we obtained in Sect.\ref{Spherical saddle-point} were only a local minimum of the action and there were another deeper minimum for a non-spherical density field $\dL'$ of the same order this would show up in our formulation and we could take into account the contribution of this second minimum. By contrast, the perturbative approach would not provide this second minimum (the only hint of its existence would be that the perturbative series diverges, but this is the case anyway for other reasons). Fortunately, as shown in Sect.\ref{Generating function} matters are simpler than this and in the quasi-linear regime the pdf $\cP(\dR)$ is indeed governed by this trivial spherical saddle-point.

Note however that these previous works based on the perturbative approach always used the implicit system (\ref{tau4}) for the generating function. More precisely, they used the branch of the generating function $\varphi(y)$ which runs through the origin in Fig.\ref{figphi}. As discussed in Sect.\ref{calPDF} this means that for $n<0$ they get a mere exponential cutoff $\cP(\dR) \sim e^{y_s \dR/\sigma^2(R)}$. As shown in App.\ref{A worked example} and discussed in Sect.\ref{calPDF} this is actually incorrect. Indeed, the high-density cutoff of the pdf is of the form (\ref{tailP1}) (until shell-crossing occurs) and the actual generating function $\varphi(y)$ shows a branch cut along the whole real negative axis (and not only for $y<y_s$) if we disregard shell-crossing. Thus, we see that for $n<0$ the perturbative approach fails beyond $\dR > \cG(\tau_s)$ and the high-density tail actually requires a non-perturbative treatment. In fact, the ``resummation'' of the perturbative theory at leading order performed in \cite{Ber1} yields the implicit system (\ref{tau4}) which remembers the existence of the non-perturbative unstable saddle-point $\tau_-<\tau_s$. Indeed, eq.(\ref{tau4}) can also be extended to $\tau<\tau_s$ where it yields the upper branch of $\varphi(y)$. However, in order to use the information contained in this upper branch one needs the non-perturbative method described in this article, which provides the integration contour required to take into account the contribution of this unstable saddle-point, see Fig.\ref{figCy1}, and which gives the full justification of this procedure. 

This unstable saddle-point modifies the pdf $\cP(\dR)$ for $\dR>\cG(\tau_s)$. This implies that the moments and the cumulants of the pdf are also changed. Therefore, at finite $\sigma$ they are not given by the expansion around $y=0$ as in eq.(\ref{cum1}) of the ``regularized'' generating function $\varphi(y)$ obtained from (\ref{tau4}) (as discussed in Sect.\ref{calPDF} the exact generating function $\varphi(y)$ for finite $\sigma$ is not regular at the origin). Moreover, the additional contribution to the moments of the pdf which arises from this shallower high-density cutoff is non-perturbative. Indeed, from eq.(\ref{tailP1}) we see that the change induced by this correction is of order:
\beq
\Delta \lag \dR^q \rag \sim e^{-\cG(\tau_s)^{(n+3)/3}/\sigma^2(R)}
\label{cumnonpert1}
\eeq
since the pdf is only modified for $\dR>\cG(\tau_s)$. It is clear that this correction cannot be obtained by a direct perturbative treatment, where one would derive the moments of the pdf by computing in serial order the terms which arise from a perturbative expansion of the density field, as in \cite{Goroff1}. Indeed, the correction (\ref{cumnonpert1}) cannot be written as a perturbative expansion over powers of $\sigma$ (it vanishes faster than any power of $\sigma$ in the limit $\sigma \rightarrow 0$). This is why the results of this iterative perturbative method actually coincide with the the results obtained from eq.(\ref{cum1}) by expanding the ``regularized'' generating function $\varphi(y)$ defined by eq.(\ref{tau4}). Thus, we see that a standard iterative perturbative method, as described in \cite{Goroff1}, cannot give the pdf $\cP(\dR)$ for $\dR>\cG(\tau_s)$ with $n<0$. In particular, these features imply that the perturbative series diverge. Note that these arguments do not involve shell-crossing. Thus, for $n<0$ the divergence of the perturbative series is not due to shell-crossing which comes into play at larger densities. Of course, as noticed in \cite{paper1} and \cite{paper5}, for $n \geq 0$ shell-crossing also leads to a divergence of the perturbative series. Then, we must point out that for large density contrasts ($\dR \ga 100$) shell-crossing must be taken into account as the spherical collapse solution is no longer described by eq.(\ref{Fcol1}). Thus, the results obtained in this article only apply to smaller density contrasts.

\subsection{Spherical model}
\label{Spherical model}

We can also note that the saddle-point method developed in this article is somewhat similar to the spherical model presented in \cite{Val1} (\S 2). This model is based on the ``educated guess'':
\beq
\int_{\dR}^{\infty} \d\delta \; (1+\delta) \cP(\delta) \simeq \int_{\delta_{L,R_L}}^{\infty} \d\dL \; \cP_L(\dL) .
\label{spher1}
\eeq
It merely states that the fraction of matter enclosed within spherical cells of radius $R$ and density contrast larger than $\dR$ (in the actual non-linear density field) is approximately equal to the fraction of matter which was originally enclosed within spherical cells of radius $R_L$ and overall linear density contrast larger than $\delta_{L,R_L}$. Here $R_L$ and $\delta_{L,R_L}$ are related to $R$ and $\dR$ as in eq.(\ref{F1}), using the spherical collapse solution. Note that this is very close in spirit to the Press-Schechter prescription used to estimate the mass function of just-collapsed objects (\cite{PS}). Then, from eq.(\ref{spher1}) and eq.(\ref{F1}) one obtains the expression (\ref{PdRdisc2}) for the pdf $\cP(\dR)$. In terms of the more usual variable $\nu=\delta_{L,R_L}/\sigma(R_L)$ this can also be written as eq.(\ref{Ptail2}). Hence, we see that the modified exponential tail (\ref{tailP1}) can be understood in very simple terms. It is directly related to the Gaussian cutoff of the linear density fluctuations and to the slope of their power-spectrum.

In \cite{Val1} we showed that the generating function $\varphi(y)$ of the quasi-linear regime, defined by eq.(\ref{tau4}), could be recovered from eq.(\ref{spher1}) and the calculation involved a saddle-point as in the present calculation. However, that previous work was a simple phenomenological study, based on a simplified description of the density field. By contrast, the present work is a rigorous study based on the exact equations of motion and we deal with the exact 3-dimensional density field. In particular, we do not require the density field to be spherically symmetric. We recover the results of the simple spherical model because the saddle-point is spherically symmetric and at leading order the generating function $\varphi(y)$ is given by the value of the action at this point. However, our results should differ when we consider higher-order terms.

Here we must note that, as described in Sect.\ref{calPDF} and App.\ref{Application}, for $n<0$ we need the prefactor $[\Det (\DL . M_y)]^{-1/2}$ for the high-density tail of the pdf (i.e. $\dR>\cG(\tau_s)$). We did not derive this determinant in a rigorous manner hence the multiplicative factor which appears in eq.(\ref{PdRdisc2}) may not be exact. In fact we can expect a non-zero correction to the approximation we used in App.\ref{Application} because the exact problem we investigate here shows some important differences with the simplified spherical model (\ref{spher1}). Indeed, this latter model only involves ordinary integrals and a one-dimensional variable $\delta_{L,R_L}$. By contrast, the formation of large-scale structures in the universe involves the infinite-dimensional variable $\dL(\bx)$ which leads to a path-integral formalism. Then, we can expect the integrations over the fluctuations around the saddle-point to show some differences between both cases. However, as seen in Fig.\ref{figPdelta} the expression (\ref{PdRdisc2}) should provide a reasonably good approximation to the exact high-density tail. In particular, it should be sufficient for practical purposes. In fact, it is probably even sufficient to use the pdf obtained from the ``regularized'' generating function $\varphi(y)$ (i.e. the lower branch in Fig.\ref{figphi}) through eq.(\ref{P2}). Note in any case that the exponential term obtained in eq.(\ref{PdRdisc2}) is exact, since it only depends on the value of the action $S[\dL]$ at the spherical saddle-point derived in Sect.\ref{Spherical saddle-point} and not on the second-derivative of the action.

\subsection{Velocity field}
\label{Velocity field}

We can note that using perturbative methods as in \cite{Ber2} or the approximate spherical model (\ref{spher1}), see \cite{Val1}, it is also possible to derive the pdf (and the associated generating function) of the mean divergence $\theta \equiv (\nabla . \bv)/{\dot a}$ of the peculiar velocity field $\bv$ within spherical cells. We shall not compute explicitly this pdf $\cP(\theta)$ here, using the saddle-point method we developed in the previous sections. Indeed, it is clear that we must recover the results of the hydrodynamical perturbative approach (i.e. the same generating function $\varphi_{\theta}(y)$). In fact, as long as the test-field $j(\bx,\bp,t)$ which enters the functional $Z[j]$ defined in eq.(\ref{Z1}) is spherically symmetric we can look for a spherical saddle-point. Then, since the physics involved is the same as the one which governs the behaviour of the pdf $\cP(\dR)$ we shall recover the same spherical saddle-point and the results of the hydrodynamical perturbative method, with the appropriate modification of the tail arising from large densities, as in eq.(\ref{tailP1}). Note that for the divergence $\theta$ the pdf shows an exponential tail for $n=-1$ (e.g., \cite{Val1}) so that the feature which appeared for $n<0$ (i.e., the singularity $y_s$) for $\cP(\dR)$ is now obtained for $n<-1$ for $\cP(\theta)$.

\section{Conclusion}

In this article, we have developed a non-perturbative method to obtain the pdf $\cP(\dR)$ of the density contrast within spherical cells in the quasi-linear regime. This corresponds to a rare-event limit: the rms fluctuation $\sigma$ vanishes while the density contrast is kept fixed. Then, a saddle-point approximation yields asymptotically exact results in this limit. Note that our approach does not rely on the hydrodynamical approximation for the equations of motion. It is fully consistent with the collisionless Boltzmann equation. However, it happens that the spherical saddle-point which governs the quasi-linear regime is an exact solution of both formalisms (hydrodynamics and Boltzmann equation). This makes the problem rather simple and it does not introduce any approximation. This is also consistent with the fact that the perturbative series obtained from the hydrodynamical and the Boltzmann frameworks are identical, see \cite{paper1}. Although the numerical examples described in this article were obtained for a critical density universe our method applies to any cosmological model. One simply needs to use the relevant spherical collapse solution $\cF(\dL)$ associated with the required values of the cosmological parameters $\Om$ and $\Ol$.

Thus, we have recovered most of the results obtained by the usual perturbative method for Gaussian primordial density fluctuations. This provides a rigorous justification of these results. Moreover, we have corrected an error introduced in these previous works for the high-density tail of the pdf for power-spectra with $n<0$. This clearly shows that one should not ask too much from perturbative methods, especially since all perturbative series actually diverge which gives room for strong non-perturbative corrections.

Note that our approach is actually much more intuitive and simpler than the perturbative method. In particular, the spherical collapse solution of the dynamics appears naturally in this framework as a saddle-point of the action, simply through the spherical symmetry of the problem. This symmetry is due to the homogeneity and isotropy of the primordial density fluctuations and to the fact that we consider the density contrast $\dR$ within spherical cells. Then, we have described a geometrical construction of the generating function $\varphi(y)$ (related to the Laplace transform of the pdf) which allows one to see at a glance its main features.

To conclude, we note that the approach developed in this article presents the advantage to introduce a method which is of standard use in physics. In particular, it makes the physics involved rather transparent. Finally, another advantage of our approach is that in principle it can also be applied to non-Gaussian primordial density fluctuations. This will be described in a companion paper (\cite{paper3}). Besides, since it is non-perturbative and it does not rely on the hydrodynamical description it could also be applied to the non-linear regime. In this case, it would give the tails of the pdf $\cP(\dR)$ (the saddle-point approximation only yields asymptotic results). We shall present a study of this non-linear regime in a future work, see \cite{paper4}.

\appendix

\section{A worked example: the lognormal pdf}
\label{A worked example}

Here we apply the steepest-descent method to the lognormal probability distribution function. This allows us to illustrate on a simple example the features implied by pdfs with a rare-event tail which decreases more slowly than an exponential cutoff. This also corresponds to generating functions $\psi(y)$ and $\varphi(y)$ which exhibit a branch cut on the negative real axis.

In order to facilitate the comparison with the problems dealt with in Sect.\ref{Saddle-point method} we shall use the same notations as far as possible. Thus, from a Gaussian variable $\dL$ with the pdf:
\beq
\cP(\dL) = \frac{1}{\sqrt{2\pi} \sigma} \; e^{-\dL^2/(2\sigma^2)}
\label{PdL}
\eeq
we define the new variable $\rho$:
\beq
\rho \equiv e^{\dL} \hspace{0.2cm} \mbox{hence} \hspace{0.2cm} \cP(\rho) = \frac{1}{\rho} \; \frac{1}{\sqrt{2\pi} \sigma} \; e^{- (\ln \rho)^2 /(2\sigma^2) } .
\label{Plog}
\eeq
Thus, $\dL$ is the analog of the linear density contrast while $\rho=1+\delta$ is the non-linear overdensity. In particular, $\dL$ runs from $-\infty$ to $+\infty$ while $\rho \geq 0$. Here it will be more convenient to study $\rho$ rather than $\delta$, so as to avoid unnecessary factors $-1$. The main difference with the problems studied in Sect.\ref{Saddle-point method} is that $\dL$ and $\rho$ are simple random variables and no longer random fields. Hence the path-integrals of the main text are replaced here by ordinary integrals. This simplifies the discussion and it will allow us to compare the predictions of the steepest-descent method with the exact pdf (\ref{Plog}).

The pdf (\ref{Plog}) is a lognormal law. We did not shift the mean of the Gaussian variable $\dL$ in order to ensure that $\lag\rho\rag=1$ since it is irrelevant for our illustrative purposes. Moreover, we do have $\lag\rho\rag=1$ at the leading order in the limit $\sigma \rightarrow 0$. The moments of the pdf $\cP(\rho)$ can be easily computed from eq.(\ref{PdL}) which yields:
\beq
\lag \rho^q\rag = \lag e^{q \dL} \rag = e^{q^2 \sigma^2/2} .
\label{momentlog}
\eeq
As in eq.(\ref{psi1}) we can define the generating function:
\beq
\psi(y) \equiv \lag e^{-y \rho} \rag = \int_0^{\infty} \d\rho \; e^{-y \rho} \; \cP(\rho)  .
\label{psilog1}
\eeq
Making the change of variable $\rho \rightarrow \dL$ we get:
\beq
\psi(y) = \int_{-\infty}^{\infty} \d\dL \; \frac{1}{\sqrt{2\pi} \sigma} \; e^{- y \; e^{\dL} -\dL^2/(2\sigma^2)} .
\label{psilog2}
\eeq
This expression clearly shows that $\psi(y)$ is not well-defined for negative real $y$ since in this case the integral in eq.(\ref{psilog2}) diverges for large positive $\dL$. This also implies that the series expansion of $\psi(y)$ diverges for all non-zero $y$:
\beq
\psi(y) = \sum_{q=0}^{\infty} \frac{(-y)^q}{q!} \; \lag \rho^q \rag = \sum_{q=0}^{\infty} \frac{(-y)^q}{q!} \; e^{q^2 \sigma^2/2} .
\label{serieslog}
\eeq

Next, let us define as in eq.(\ref{psisig}) the rescaled generating functions $\varphi(y)$ and $\psib(y)$:
\beq
\psi(y) \equiv \psib(y \sigma^2) \equiv e^{-\varphi(y \sigma^2)/\sigma^2} .
\label{philog1} 
\eeq
As in eq.(\ref{tau2}) and eq.(\ref{G1}) we also introduce the variable $\tau$ and the function $\cG(\tau)$ by:
\beq
\tau \equiv - \dL  , \hspace{0.2cm} \cG(\tau) \equiv e^{-\tau} = \rho .
\label{Glog1}
\eeq
Then, we can write eq.(\ref{psilog2}) as:
\beq
\psib(y) = \int_{-\infty}^{\infty} \frac{\d\tau}{\sqrt{2\pi} \sigma} \; e^{- [ y \; \cG(\tau) + \tau^2/2 ]/\sigma^2} .
\label{psilog3}
\eeq
This expression is the analog of the path-integral (\ref{phi1}). In particular, a steepest-descent method yields again exact results in the limit $\sigma \rightarrow 0$. Moreover, the saddle-point and the value of the generating function $\varphi(y)$ are again given by eq.(\ref{tau4}). Then, from $\varphi(y)$ one obtains the pdf $\cP(\rho)$ (in the quasi-linear limit) through the inverse Laplace transform (\ref{P2}).

This steepest-descent method is fully justified for positive real $y$ where the integral (\ref{psilog3}) converges. Since negative $\delta$ (i.e. $\rho \leq 1$) corresponds to positive $\tau_c$ and $y_c$ this method yields the pdf $\cP(\rho)$ for $\rho \leq 1$. However, for negative real $y$ the integral diverges. Hence one cannot directly apply this procedure for $\rho > 1$ since the saddle-point $y_c$ which would appear in the computation of eq.(\ref{P2}) would be negative. Nevertheless, the steepest-descent method is still useful but it must be applied with some care, as we shall describe below. A similar problem arises in usual Quantum Field Theory when one tries to derive non-perturbative results from path-integrals. This leads to the so-called ``instanton'' contributions, see \cite{Zinn1}. However, since some features are specific to our case (e.g., the saddle-points $\tau_c$ are not fixed) we shall detail the procedure required by the problem we investigate.

\begin{figure}[htb]
\centerline{\epsfxsize=8cm \epsfysize=4.5cm \epsfbox{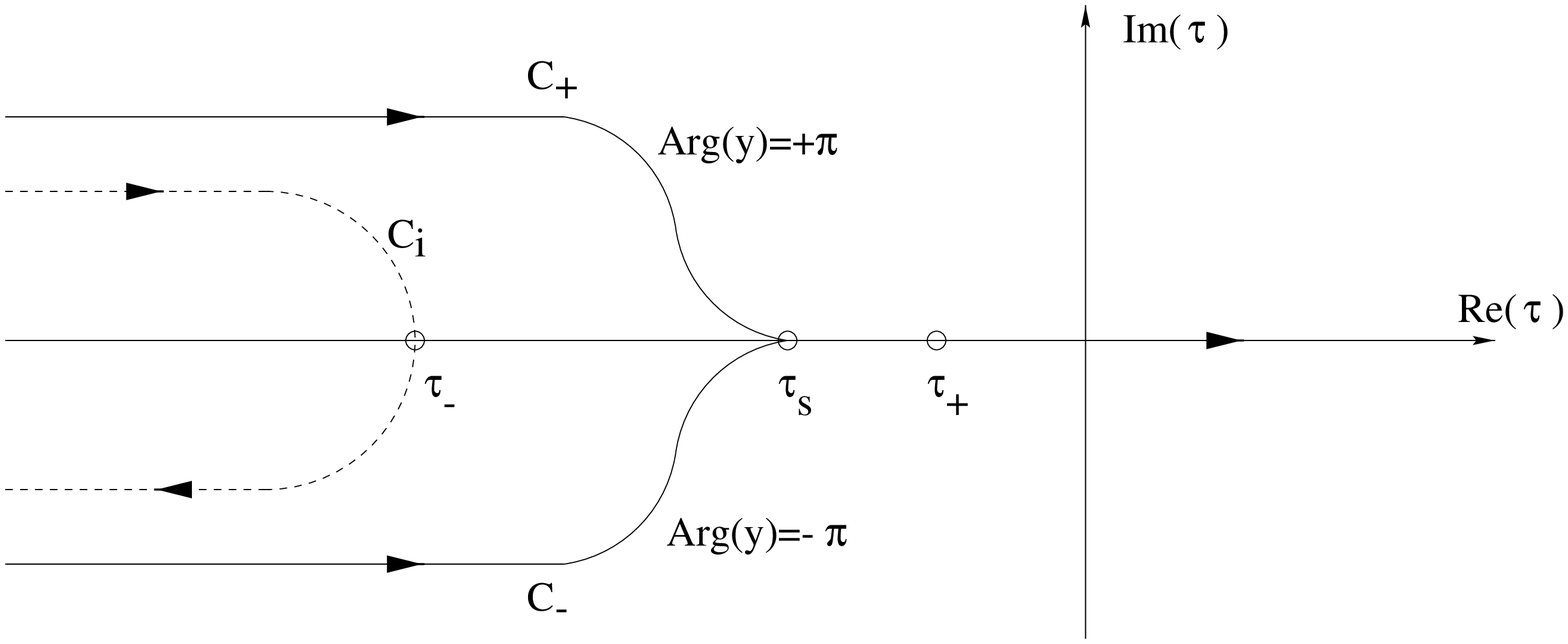}}
\caption{The integration path $\cC$ over $\tau$ in the complex plane. The paths $\Cp$ and $\Cm$ shown in the figure correspond to $\Arg(y)=\pi$ and $\Arg(y)=-\pi$. They run from $\Re(\tau)=-\infty$ (where $\Im(\tau) = \pm \pi$) up to $\Re(\tau)=+\infty$ (where $\Im(\tau) = 0$). They give the integration path for the generating function $\psib(y)$. The dashed-line shows the contour $\Ci$ which yields the discontinuity of $\psib(y)$ along the real negative axis (branch cut). The saddle-points $\tau_-$ and $\tau_+$ on the real axis correspond to small $y$: $y_s < y <0$.} 
\label{figC1}
\end{figure}

First, we need to perform the analytic continuation of $\psib(y)$ over the complex plane, starting from real positive $y$. To do so, we must deform the integration path $\cC$ in eq.(\ref{psilog3}) as we change the argument of $y$ so that $\Re(S)$ remains positive for $\tau \rightarrow -\infty$, where the ``action'' $S$ is:
\beq
S \equiv y \; \cG(\tau) + \frac{\tau^2}{2} = y \; e^{-\tau} + \frac{\tau^2}{2}.
\label{Slog1}
\eeq
Since this deformation must be continuous as we increase (or decrease) $\Arg(y)$ from the case $\Arg(y)=0$ (where $\cC$ is the real axis) one can easily check that the contour $\cC$ is of the form shown in Fig.\ref{figC1}. It obeys:
\beq
\Re(\tau) \rightarrow -\infty : \;\; \Im(\tau) = \Arg(y) .
\eeq
For $\Re(\tau) \rightarrow +\infty$ we can keep $\Im(\tau)=0$ (actually, we only need $|\Arg(\tau)|<\pi/4$). Thus, for $y = - |y| + i 0$ we have the contour $\Cp$ with $\Im(\tau) = \pi$ for large negative $\Re(\tau)$ while for $y = - |y| - i 0$ we have the contour $\Cm$ with $\Im(\tau) = -\pi$. Note that $\psib(y+i0) \neq \psib(y-i0)$, where $y$ is real negative.

We also display in Fig.\ref{figC1} the points $\tau_-$, $\tau_s$ and $\tau_+$ obtained for a small negative $y$. As in Sect.\ref{Geometrical construction}, the point $\tau_+$ is a local minimum of the action $S$ along the real axis. In fact, it is the global minimum along the integration path $\cC$. On the other hand, $\tau_-$ is a local maximum along the real axis. This implies that the steepest-descent path runs through $\tau_+$ along the real axis while it runs through $\tau_-$ perpendicularly to the real axis (so that $\tau_-$ is a local minimum). From eq.(\ref{sing1}) the singular point $(\tau_s,y_s)$ is given by:
\beq
\tau_s = -1 , \;\; y_s = -1/e .
\label{singlog1}
\eeq
Let us first consider for small $y<0$ the sum of both contributions:
\beqa
2 \; \Re[ \psib(y) ] & \equiv & \psib(y+i0) + \psib(y-i0) \nonumber \\ & = & \int_{\Cp+\Cm} \frac{\d\tau}{\sqrt{2\pi} \sigma} \; e^{- [ y \; \cG(\tau) + \tau^2/2 ]/\sigma^2}
\label{psilog4}
\eeqa
where we note $\psib(y) = \psib(y+i0)$ for $y<0$. Then, we see from Fig.\ref{figC1} that the saddle-point $\tau_+$ contributes to $\Re[ \psib(y) ]$ since both contours $\Cp$ and $\Cm$ run through this point along the real axis in the same direction (their contributions are equal and they sum up). On the contrary, the saddle-point $\tau_-$ gives no contribution. Indeed, even if we deform the contours $\Cp$ and $\Cm$ so that they run vertically through $\tau_-$ in Fig.\ref{figC1} their contributions are of opposite sign (both contours are symmetric with respect to the real axis). Next, the discontinuity of $\psib(y)$ along the branch cut (i.e. the real negative axis) is given by the difference between both integrals:
\beqa
2 i \; \Im[ \psib(y) ] & \equiv & \psib(y+i0) - \psib(y-i0) \nonumber \\ & = & \int_{\Ci} \frac{\d\tau}{\sqrt{2\pi} \sigma} \; e^{- [ y \; \cG(\tau) + \tau^2/2 ]/\sigma^2} .
\label{psilog5}
\eeqa
Indeed, the contour $\Cp-\Cm$ can be deformed into the contour $\Ci$ shown by the dashed line in Fig.\ref{figC1}. Thus, the discontinuity is governed by the saddle-point $\tau_-$ and the contribution from $\tau_+$ cancels. Note however that this behaviour only applies to small negative $y$, $y_s<y<0$, where the two real saddle-points $\tau_-$ and $\tau_+$ exist. For $y=y_s$ these two saddle-points merge and for $y<y_s$ we have two complex conjugate saddle-points. However, we shall not need study this regime.

\begin{figure}[htb]
\centerline{\epsfxsize=8cm \epsfysize=4.5cm \epsfbox{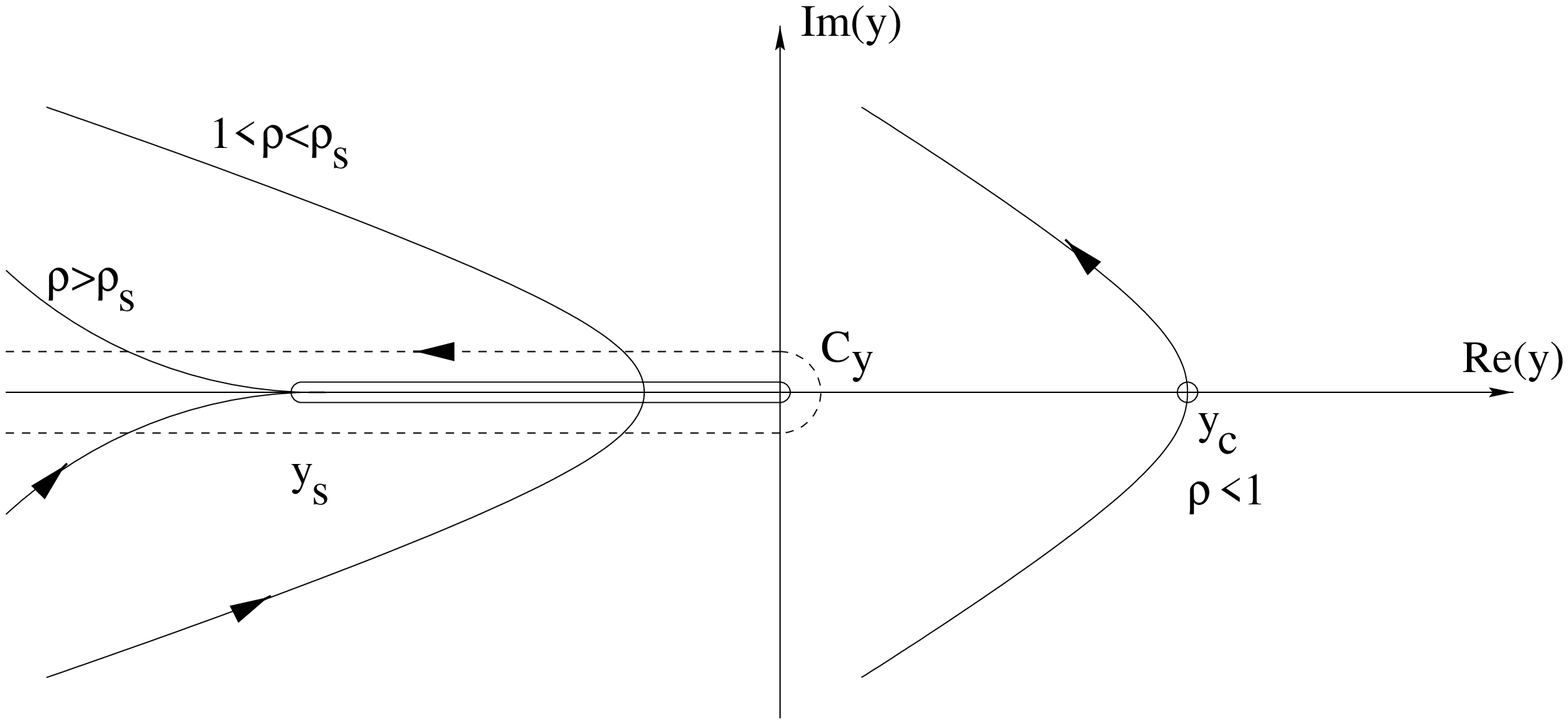}}
\caption{The integration path $\cC_y$ (dashed curve) over $y$ in the complex plane. For $\rho<1$ this integration path can be deformed into the contour shown by the solid curve on the right which runs through the saddle-point $y_c$ on the real axis. For $1<\rho<\rho_s$ or $\rho>\rho_s$ we use the two contours shown by the solid curves on the left (see main text).} 
\label{figCy1}
\end{figure}

The pdf $\cP(\rho)$ is obtained from the generating functions $\psib(y)$ or $\varphi(y)$ through the inverse Laplace transform as in eq.(\ref{P2}). This now reads:
\beq
\cP(\rho) = \int_{\Cy} \frac{\d y}{2 \pi i \sigma^2} \; e^{\rho y/\sigma^2} \; \psib(y)
\label{Plog2}
\eeq
Since $\psib(y)$ has a branch cut along the real negative axis the contour $\Cy$ bends around the real negative axis and it intersects the real axis at $y>0$. The integration path $\Cy$ is shown by the dashed curve in Fig.\ref{figCy1}. As we have seen above, the saddle-point $\tau_+$ gives the dominant contribution to $\Re(\psib)$ for small $y$. If we only take into account this contribution, we obtain the generating function $\varphi(y)$ defined by eq.(\ref{tau4}). This corresponds to the branch for $y>y_s$ which runs through the origin in Fig.\ref{figphi}. Note that this function $\varphi(y)$ is no longer singular at $y=0$. Indeed, as shown above the discontinuity of $\psib(y)$ along the branch cut (which starts at $y=0^-$) is given by the saddle-point $\tau_-$ and not by $\tau_+$, see eq.(\ref{psilog4}) and eq.(\ref{psilog5}). This ``regularized'' function $\varphi(y)$ still exhibits a branch cut for $y<y_s$ when the saddle-point $\tau_+$ gives rise to two complex conjugate saddle-points. Of course, since these generating functions $\varphi(y)$ and $\psib(y)$ are regular at the origin, their expansion (\ref{serieslog}) now converges for $|y| < |y_s|$. Note indeed that the moments of the pdf obtained in this way are no longer given by eq.(\ref{momentlog}) since we only keep the leading order of the cumulants in the limit $\sigma \rightarrow 0$. This describes the pdf for small positive density contrasts $\delta$ ($\rho \sim 1$). Moreover, one can check by a direct calculation from eq.(\ref{momentlog}) that the expansion around $y=0$ of the ``regularized'' function $\varphi(y)$ defined by eq.(\ref{tau4}) indeed yields the cumulants of the pdf $\cP(\rho)$ at the leading order in $\sigma \rightarrow 0$.

However, the behaviour of the pdf $\cP(\rho)$ for large $\rho$ is not governed by $\tau_+$ but by the saddle-point $\tau_-$. First, let us note that the pdf obtained by this ``regularized'' function $\varphi(y)$ exhibits an exponential cutoff of the form $\cP(\rho) \sim e^{y_s \rho/\sigma^2}$. This is obvious from eq.(\ref{Plog2}). Indeed, since these ``regularized'' generating functions $\varphi(y)$ and $\psib(y)$ show a branch cut for $y<y_s$ the integration path over $y$ in eq.(\ref{Plog2}) is bent around this branch cut and for large $\rho$ (or $\delta$) the integral is dominated by $y \simeq y_s$ since the other parts of the path with $\Re(y) < y_s$ become exponentially small as $e^{y \rho/\sigma^2}$ with respect to $e^{y_s \rho/\sigma^2}$ (e.g., \cite{Ber1}). Second, as we have noticed above the exact generating functions $\psib(y)$ and $\varphi(y)$ actually show a branch cut along the real negative axis for $y<0$. Then, it is clear that for large $\rho$ the pdf is governed by the singularity at $y=0$. Indeed, formally this leads to a cutoff $\cP(\rho) \sim e^{y_s' \rho/\sigma^2}$ with $y_s' \rightarrow 0$ for large $\rho$. This actually corresponds to a pdf with a large density tail which decreases more slowly than an exponential. Of course, we can check that this agrees with eq.(\ref{Plog}).

This property can be derived as follows from eq.(\ref{Plog2}). If we make the upper and lower branches to get very close to the real negative axis we can use eq.(\ref{psilog5}) to write:
\beq
\cP(\rho) = - \int_{-\infty}^0 \frac{\d y}{\pi \sigma^2} \; e^{\rho y/\sigma^2} \; \Im [ \psib(y) ] .
\label{Plog3}
\eeq
Thus we integrate the discontinuity along the real negative axis. Then, in the limit $\rho \rightarrow \infty$ the integral (\ref{Plog3}) is dominated by the behaviour of $\Im [ \psib(y) ]$ in the neighbourhood of $y \simeq 0^-$. As was shown in eq.(\ref{psilog5}), in this regime $\Im [ \psib(y) ]$ is governed by the saddle-point $\tau_-$. Indeed, for $y \rightarrow 0^-$ we have $y_s<y<0$ so that the saddle-points $\tau_-$ and $\tau_+$ lie on the real axis. Note that the limit $\rho \rightarrow \infty$ is not the limit $\sigma \rightarrow 0$. In other words, in order to obtain the high-density tail of the pdf $\cP(\rho)$ one must keep $\sigma$ finite (even though small) and study the limit of large $\rho$.

We shall now derive the high-density tail of the pdf $\cP(\rho)$. First, as explained above we need $\Im [ \psib(y) ]$ for $y \rightarrow 0^-$. This is given by the saddle-point $\tau_-$ in eq.(\ref{psilog5}). Thus, a Gaussian integration yields:
\beq
2 i \; \Im[ \psib(y) ] = -i \left[ - 1 - y \cG''(\tau_-) \right]^{-1/2} \; e^{-S_y/\sigma^2}
\label{Impsi1}
\eeq
where $S_y \equiv S(\tau_-)$ is given in eq.(\ref{Slog1}). Then, as in the discussion of eq.(\ref{Chic1}) the pdf $\cP(\rho)$ in eq.(\ref{Plog3}) is given by an integration around the saddle-point $(y_c,\tau_c)$. However, the integration path now runs along the real negative axis and we have $y_s<y_c<0$ and $\tau_c<\tau_s$. It is interesting to consider the steepest-descent approximation for this integral too. This yields:
\beq
\cP(\rho) = \frac{1}{\sqrt{2\pi}\sigma} \; \frac{1}{\sqrt{-1-y\cG''(\tau)}} \; \frac{1}{\sqrt{\varphi''(y)}} \; e^{-\tau^2/(2\sigma^2)}
\label{Plog4}
\eeq
where $\tau$ obeys $\cG(\tau) = \rho$ while $y$ and $\varphi(y)$ are given by the implicit system (\ref{tau4}). More precisely, $\varphi(y)$ is now described by the upper branch in Fig.\ref{figphi} which runs over $y_s<y<0$. From eq.(\ref{tau4}) we also obtain:
\beq
\varphi''(y) = \cG'(\tau) \frac{\d\tau}{\d y} , \;\;\; -1-y\cG''(\tau) = \cG'(\tau) \frac{\d y}{\d\tau} .
\eeq
This yields:
\beq
\cP(\rho) = \frac{1}{\sqrt{2\pi}\sigma} \; \frac{1}{|\cG'(\tau)|} \; e^{-\tau^2/(2\sigma^2)} .
\label{Plog5}
\eeq
Thus, we actually recover the exact pdf (\ref{Plog}) since from eq.(\ref{Glog1}) we have: $\tau=-\ln\rho$ and $\cG'(\tau)=-\rho$. This is natural for large $\rho$ since the steepest-descent method described above becomes exact in this limit. For smaller $\rho$ the pdf given by eq.(\ref{Plog2}), where we perform the exact numerical integration, is no longer equal to the exact pdf (\ref{Plog}). This is the generic case. The pdf obtained by the steepest-descent method only yields asymptotically exact results in the two limits $\sigma \rightarrow 0$ or $\rho \rightarrow \infty$.

\begin{figure}[htb]
\centerline{\epsfxsize=8cm \epsfysize=5.8cm \epsfbox{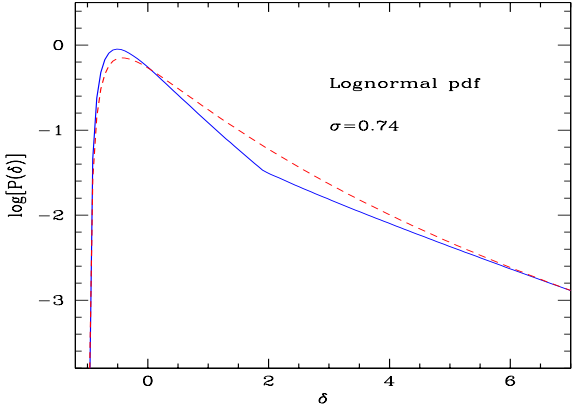}}
\caption{The pdf $\cP(\delta)$ for the lognormal case with $\sigma=0.74$. The dashed-curve shows the exact pdf from eq.(\ref{Plog}). The solid curve shows the results obtained by the steepest-descent method. The break at $\delta \sim 2$ corresponds to the transition to the high-density tail which is governed by the saddle-point $\tau_-$.} 
\label{figPdeltalog}
\end{figure}

Finally, using the steepest-descent method described above we can estimate the pdf $\cP(\rho)$. As discussed above, the exact generating functions $\psib(y)$ and $\varphi(y)$ show a branch cut on the real negative axis so the integration path over $y$ in eq.(\ref{Plog2}) is bent around this cut as shown by the dashed curve in Fig.\ref{figCy1}. For $\rho<1$ this contour can be deformed in the path shown by the right solid curve which runs through the saddle-point $y_c>0$ on the real axis. For $\rho>1$ matters are more intricate as discussed above. For $1<\rho<\rho_s$, where $\rho_s=\cG(\tau_s)=1/e$, we only take into account the ``regularized'' part of $\varphi(y)$ which is described by eq.(\ref{tau4}). This function is regular over $y>y_s$ and we use the integration path shown by the solid curve labeled ``$1<\rho<\rho_s$'', which runs through a saddle-point $y_s<y_c<0$ on the real axis. Next, for large densities $\rho>\rho_s$ we take into account the exact branch cut along the real axis in order to obtain the correct high density tail. This is shown by the contour labeled ``$\rho>\rho_s$''. We split this path into two parts. The first one around the real negative axis with $y_s\leq y\leq0$ is governed by the saddle-point $\tau_-$ which leads to a saddle-point $y_c$ which goes to $0^-$ for large $\rho$. The second part over the range $-\infty < \Re(y)<y_s$ is computed using eq.(\ref{tau4}) for $\varphi(y)$. Then, we simply take the largest of these two contributions to estimate $\cP(\rho)$. In fact, for $\rho>\rho_s$ the first part is the largest one in the limit $\sigma \rightarrow 0$, as expected. For $\sigma \rightarrow 0$ or $\rho \rightarrow \infty$ it is actually infinitely larger.

We display in Fig.\ref{figPdeltalog} our results for the pdf $\cP(\delta)$ with $\sigma=0.74$ (we show $\cP(\delta)$ rather than $\cP(\rho)$ in order to compare with the main section). The results obtained by the steepest-descent method are shown by the solid curve. The knee at $\delta \sim 2$ corresponds to the transition from the ``regularized'' $\varphi(y)$ to the exact branch cut at $y<0$. Thus, for $\delta \ga 2$ the pdf is governed by the saddle-point $\tau_-$ and the neighbourhood of $y=0^-$. We can check in the figure that this estimate is indeed exact in the limit $\rho \rightarrow \infty$, as we proved in eq.(\ref{Plog5}). For smaller $\rho$ the pdf is governed by the saddle-point $\tau_+$, that is $\varphi(y)$ is described by the branch which runs through the origin given by eq.(\ref{tau4}). Thus, we see that the steepest-descent method provides reasonably good results up to $\sigma=0.74$. However, the agreement is not as good as in Fig.\ref{figPdelta} for the actual pdf $\cP(\dR)$ which arises from gravitational clustering.

\section{Application to path-integrals}
\label{Application}

We have described in App.\ref{A worked example} how to apply the steepest-descent method in the case where the function $\cG(\tau)$ grows faster than $\tau^2$ for $\tau \rightarrow -\infty$. As noticed in Sect.\ref{Geometrical construction} and Sect.\ref{calPDF}, this corresponds to linear power-spectra $P(k)$ with $n<0$ for the problem of gravitational clustering which we investigate in this article. The problem studied in App.\ref{A worked example} actually involved ordinary one-dimensional integrals but the arguments can be generalized to path-integrals. Note that in our case the function $\cG(\tau)$ does not grow as $e^{-\tau}$ (as in App.\ref{A worked example}) but as a power-law, see eq.(\ref{tauG1}). Then, the contour $\cC$ in the complex plane over $\tau$ which was shown in Fig.\ref{figC1} is now given by:
\beq
\Re(\tau) \rightarrow -\infty : \;\; \Arg(\tau) = \pi - \frac{n+3}{6} \Arg(y).
\label{Ctaun1}
\eeq
However, this does not change the behaviour described in App.\ref{A worked example}.

Thus, we can directly apply to the path-integral (\ref{phi1}) the procedure detailed in App.\ref{A worked example}. However, for $n<0$ where the high-density tail is governed by the saddle-point $\tau_-$, the transposition of the Gaussian integration of eq.(\ref{psilog5}) which yielded eq.(\ref{Impsi1}) now gives a factor:
\beq
2 i \; \Im[ \psib(y) ] = \left( \Det\DL^{-1} \right)^{1/2} \left( \Det M_y \right)^{-1/2} e^{-S_y/\sigma^2(R)}
\eeq
as in eq.(\ref{phi2}). Note that since the spherical saddle-point $\dL(\bx)$ (i.e. $\tau_-$) is now a local maximum of the action $S[\dL]$ the determinant $\Det M_y$ is negative so that the square root $\left( \Det M_y \right)^{-1/2}$ yields a factor $i$. Thus, the numerical factor which appeared in eq.(\ref{Impsi1}) is replaced by a determinant:
\beq
1 + y \cG''(\tau_-) \rightarrow \cD \hspace{0.2cm} \mbox{with} \hspace{0.2cm} \cD \equiv \Det( \DL . M_y ) .
\label{D1}
\eeq
In order to compute the pdf $\cP(\dR)$ we only need to evaluate this determinant $\cD$. Indeed, the minimum $S_y$ of the action is still given by eq.(\ref{tau4}) where we use the upper branch defined over the range $y_s<y<0$ shown in Fig.\ref{figphi}. From eq.(\ref{phi2}) we have:
\beqa
\lefteqn{ (\DL . M_y)(\bx_1,\bx_2) = \delta_D(\bx_1-\bx_2) } \nonumber \\ & & + \frac{y}{\sigma^2(R)} \int \d\bx \; \DL(\bx_1,\bx) \; \left. \frac{\delta^2 (\dR)}{\delta(\dL(\bx)) \delta(\dL(\bx_2))} \right|_{\tau}
\label{D2}
\eeqa
Thus, we need the second-order derivative of the functional $\dR[\dL]$ taken at the spherical saddle-point characterized by $\delta_{L,R_L}$ (or $\tau$). Since we do not know the explicit form of the functional $\dR[\dL]$ we shall only obtain a simple estimate of the determinant $\cD$. However, as we can see in Fig.\ref{figPdelta} the behaviour of the discontinuity of the generating functions $\psib(y)$ and $\varphi(y)$ near $y=0^-$ only dominates the pdf $\cP(\dR)$ for rather large density contrasts ($\delta \ga 8$) for $\sigma = 0.74$. Hence we do not need the high-density tail of the pdf with a high accuracy if we study the quasi-linear regime.

In order to estimate the determinant $\cD$ we could try to use second-order perturbation theory. Indeed, the second-order derivative in eq.(\ref{D2}) becomes negligible on very large scales $x \gg R_L, x_2 \gg R_L$. Moreover, the radial profile of the spherical saddle-point is almost flat in the inner region $R'<R_L$. Hence we might estimate this second-order derivative, which should be taken at the point $\dL(\bx)$ given by eq.(\ref{col5}), by its value at the point $\dL(\bx) = \delta_{L,R_L}$ (i.e. constant density contrast). Then, we simply need to investigate the second-order perturbation theory in a background universe characterized by a higher mean density: $\rhob \rightarrow \rhob (1+\dR)$. Here $\dR = \cG(\tau) = \cF(\delta_{L,R_L})$ is the actual non-linear density contrast of the spherical saddle-point. Unfortunately, this procedure cannot give meaningful results. Indeed, it is well known that perturbation theory leads to divergent quantities when one goes beyond leading order terms (e.g., \cite{Scoc1}). Then, it is easy to check that the calculation of the determinant $\cD$ from eq.(\ref{D2}) with the use of the second-order term $\delta^{(2)}(\bx)$ for the density field $\delta(\bx)$ (written as an expansion over $\dL(\bx)$) gives rise to such divergences. In fact, as shown in \cite{paper5} we can check that using a perturbative approach to evaluate the fluctuations of the action $S[\dL]$ around the saddle-point we exactly recover the divergences obtained from standard direct perturbative methods.

As a consequence, we shall use the following approximation for the generating function $\psib(y)$. By analogy with the case of ordinary integrals studied in App.\ref{A worked example} we replace in a first step the determinant $\cD$ by a factor $1 + y \cG''(\tau_-)$, see eq.(\ref{D1}). This takes into account the dependence of the non-linear density contrast on the local linear density contrast $\delta_{L,R_L}$. However, a new physical process which did not appear in App.\ref{A worked example} occurs in the context of cosmology: the expansion of the background universe. This leads to a dilution of the high-density tail of the probability distribution $\cP(\dR)$. Indeed, let us consider for a moment the following local model. At a time $t_1$ we mark the comoving coordinates $\bx$ where the linear density contrast $\delta_{L,R_{L1}}(\bx)$ over the cell $V_{L1}$ centered on $\bx$ is above some threshold $\delta_{c1}$, which corresponds to a non-linear density contrast $\Delta_{c1} = \cF(\delta_{c1})$. This fills a fraction $F_1$ of the volume of the universe. At a later time $t_2$, the same fraction $F_1$ of the universe now shows density contrasts above $\delta_{c2}=\delta_{c1} \Dgp(t_2)/\Dgp(t_1)$ and $\Delta_{c2} = \cF(\delta_{c2})$, where $\Dgp(t)$ is the linear growing mode. However, it happens that in fact the regions over the non-linear threshold $\Delta_{c2}$ no longer fill a fraction $F_1$ of the universe. Indeed, while their density increases these regions also depart from the mean background expansion and they actually contract in comoving coordinates (for positive density contrasts). Thus, we have $F_2=F_1 (R_2/R_1)^3 = F_1 (1+\Delta_{c1})/(1+\Delta_{c2})$. Therefore, we add a dilution factor $1/(1+\dR)$ to the generating function $\psib(y)$. Hence we write:
\beq
2 i \; \Im[ \psib(y) ] \simeq \frac{1}{1+\dR} \; \frac{1}{\sqrt{1+y\cG''(\tau)}} \; e^{-S_y/\sigma^2(R)}
\label{psib1}
\eeq
where $\dR=\cG(\tau)$ and $\tau(y)$ is given by the upper branch in Fig.\ref{figphi}. Next, the pdf $\cP(\dR)$ defined by the inverse Laplace transform (\ref{P1}) can be written as in eq.(\ref{Plog3}) in terms of $\psib(y)$. This yields:
\beq
\cP(\dR) \simeq \int_{y_s}^0 \frac{\d y}{2\pi \sigma^2} \frac{1}{1+\dR} \frac{1}{\sqrt{-1-y\cG''(\tau)}} e^{[y\dR-S_y]/\sigma^2} .
\label{psib2}
\eeq
Then, in the limit $\sigma \rightarrow 0$ we can evaluate this integral by an ordinary steepest-descent method, as in eq.(\ref{Plog4}). Thus we obtain:
\beq
\cP(\dR) \simeq \frac{1}{\sqrt{2\pi}\sigma} \; \frac{1}{1+\dR} \; \frac{1}{|\cG'(\tau)|} \; e^{-\tau^2/(2\sigma^2(R))} .
\label{Ptail1}
\eeq
The expression (\ref{Ptail1}) can actually be recovered from a very simple spherical model detailed in \cite{Val1}. Thus, let us define the variable $\nu$ which describes linear density fluctuation by:
\beq
\nu \equiv \frac{\delta_{L,R_L}}{\sigma(R_L)} = - \frac{\tau}{\sigma(R)} .
\label{nu1}
\eeq
Then, the expression (\ref{Ptail1}) can also be written:
\beq
\cP(\dR) \simeq \frac{1}{\sqrt{2\pi}} \; \frac{1}{1+\dR} \; \frac{\d\nu}{\d\dR} \; e^{-\nu^2/2}
\label{Ptail2}
\eeq
where we used: $\cG'(\tau) = \d\dR/\d\tau$. The relation (\ref{Ptail2}) agrees exactly with the spherical model studied in \cite{Val1}, which we discuss here in Sect.\ref{Spherical model}. This model is similar to the Press-Schechter approach (\cite{PS}) used to estimate the mass function of collapsed halos (without the factor 2). Then, eq.(\ref{Ptail2}) should be sufficient for practical purposes, as seen in Fig.\ref{figPdelta}. Note that although we did not obtain a rigorous derivation of the multiplicative factor in eq.(\ref{Ptail2}) the exponential cutoff is exact.

\end{document}